\documentclass[12pt,preprint]{aastex}
\title {The U-band Galaxy Luminosity Function of Nearby Clusters}
\author {Daniel Christlein\altaffilmark{1}, Daniel H. McIntosh\altaffilmark{2}, Ann I. Zabludoff\altaffilmark{1}}
\altaffiltext{1} {Steward Observatory, The University of Arizona, 933 N Cherry Ave, Tucson 85721 AZ}
\email {dchristlein@as.arizona.edu \\ dmac@hamerkop.astro.umass.edu \\ azabludoff@as.arizona.edu}
\altaffiltext{2} {Astronomy Department, University of Massachusetts, Amherst, MA 01003}
\begin {document}
\shortauthors{Christlein, McIntosh \& Zabludoff}
\shorttitle{U-band Galaxy Luminosity Function} 
\begin {abstract}

Despite the great potential of the $U$-band galaxy luminosity function (GLF) to constrain the history of star formation in clusters, to clarify the question of variations of the GLF across filter bands, to provide a baseline for comparisons to high-redshift studies of the cluster GLF, and to estimate the contribution of bound systems of galaxies to the extragalactic near-UV background, determinations have so far been hampered by the generally low efficiency of detectors in the $U$-band and by the difficulty of constructing both deep and wide surveys. In this paper, we present $U$-band GLFs of three nearby, rich clusters to a limit of $M_{U}\approx-17.5$ ($M^{*}_{U}+2$). Our analysis is based on a combination of separate spectroscopic and $R$-band and $U$-band photometric surveys. For this purpose, we have developed a new maximum-likelihood algorithm for calculating the luminosity function that is particularly useful for reconstructing the galaxy distribution function in multi-dimensional spaces (e.g., the number of galaxies as a simultaneous function of luminosity in different filter bands, surface brightness, star formation rate, morphology, etc.), because it requires no prior assumptions as to the shape of the distribution function. 

The composite luminosity function can be described by a Schechter function with characteristic magnitude $M_{U}^{*}=-19.82\pm0.27$ and faint end slope $\alpha_{U}=-1.09\pm0.18$. The total $U$-band GLF is slightly steeper than the $R$-band GLF, indicating that cluster galaxies are bluer at fainter magnitudes. Quiescent galaxies dominate the cumulative $U$-band flux for $M_{U}<-14$. The contribution of galaxies in nearby clusters to the $U$-band extragalactic background is $<1\%$ Gyr$^{-1}$ for clusters of masses $\sim3\times 10^{14}$ to $2\times10^{15}$ $M_{\odot}$.

 \end{abstract}

\section{INTRODUCTION}

The galaxy luminosity function (GLF) is one of the most fundamental statistics of galaxy populations. Its shape and variation with environment provide a crucial constraint on any model of galaxy evolution. Recent studies of the GLF in clusters of galaxies, based on spectroscopic surveys, have been carried out in the $R$-band \citep{cz2003} and the $b_{J}$-band \citep{depropris}. However, there is a large amount of additional information to be gleaned from $U$-band GLFs of cluster galaxies.

Why do existing luminosity functions in the $R$- and $b_{J}$-band not provide us with a complete picture of the galaxy population in clusters? Different filter bands are sensitive to different stellar populations. For example, blue and near-UV bands are most sensitive to recent star formation, while red and near-IR bands better approximate the total stellar mass. Determinations of the GLF in different magnitude bands are therefore complementary, and the $U$-band in particular promises insight into a number of important questions:

First, several studies \citep{bromley98,madgwick02,cz2003} show that star-forming and quiescent galaxies have very different GLFs. It is therefore reasonable to expect that star formation would affect the shape of the overall GLF, and that it would do so differently in different filter bands. Other phenomena, such as the presence of dust or the metallicity of a galaxy population, also affect galaxy colors and could introduce inhomogeneities in comparisons between different filter bands. Understanding how strongly such color-dependent effects influence the GLF is crucial to evaluating the significance of discrepancies between observational determinations of the GLF in different filter bands, and between observed GLFs and model predictions.

\begin{figure}
\plotone{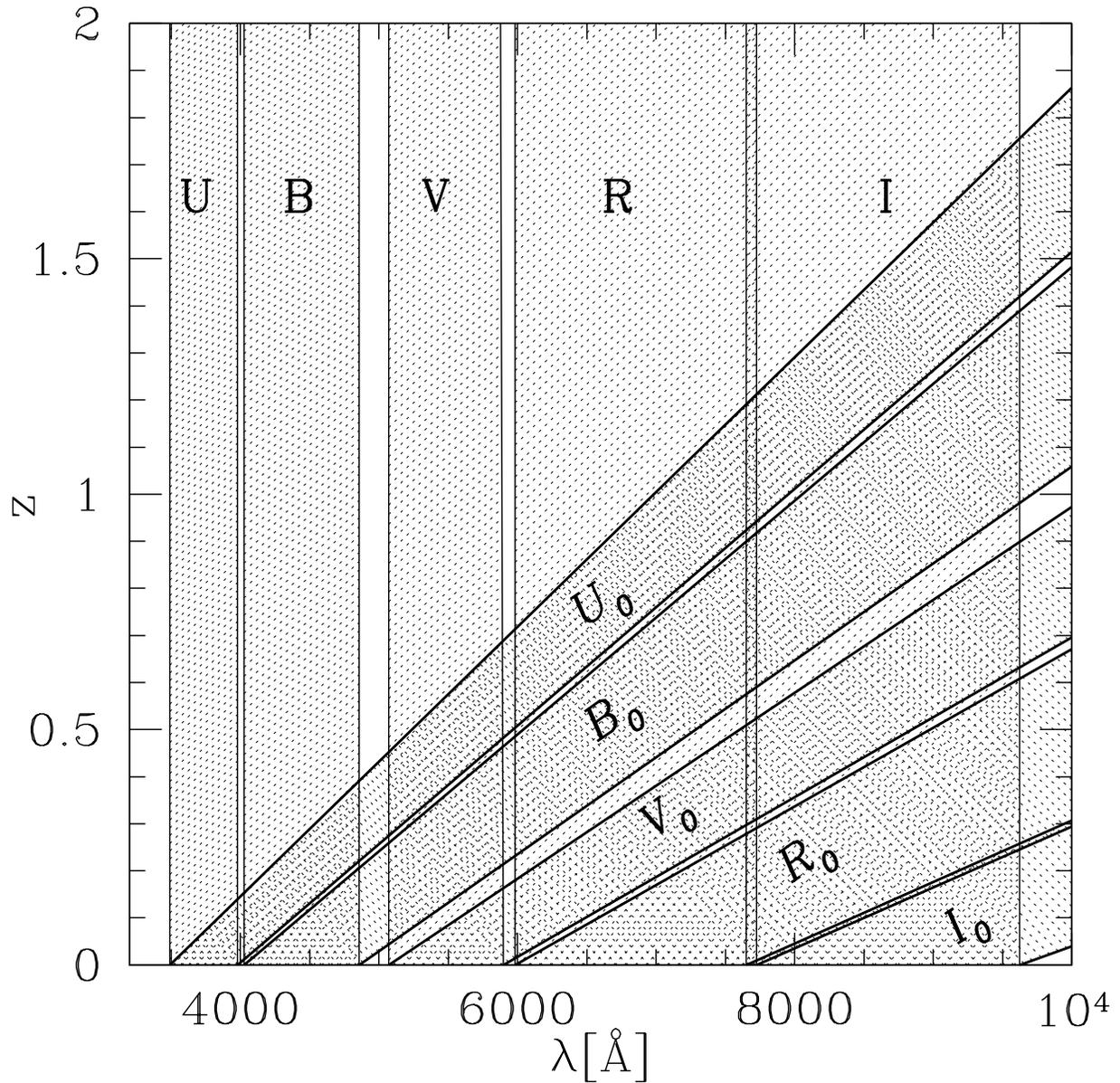}
\caption{Comparison between rest-frame UBVRI and observed U$_0$B$_0$V$_0$R$_0$I$_0$ filter bands at different redshifts. Tilted lines show how the rest-frame bands are mapped onto observed bands (vertical solid lines) for any given redshift. At z=0.8, observations in the $R$-band detect the rest-frame $U$-band.}
\label{figbands}
\end{figure}

Second, with the availability of large telescopes and new detection techniques \citep{madau}, recent years have seen growing interest in observations of clusters at high redshift. Surveys of the GLF in such systems reach to approximately 2 mag fainter than the characteristic bright magnitude $M^{*}$ at redshifts up to $z\approx1$ \citep{deproprisclusterKglf, massarotti2003}. High-redshift observations in red and near-IR bands typically observe blue or near-UV rest-frame wavelengths (Fig. \ref{figbands}). Measuring the evolution of such fundamental statistics as the GLF in these bands has been complicated by the lack of corresponding low-redshift GLFs. Our $U$-band GLF for clusters at $z\approx0$ provides a baseline for comparisons to high-redshift clusters that are observed in the $U$-band rest-frame band with a comparable depth ($<M^{*}+2$).

Third, there is an extragalactic background in the $U$-band at a level of 2 to 4$\times 10^{-9}$ erg s$^{-1}$ cm$^{-2}$ sr$^{-1}$ \AA$^{-1}$ \citep{bernstein02,henry99}. It is thought that a majority of this light is produced by stellar nucleosynthesis rather than AGN or other non-stellar sources. Therefore, the $U$-band extragalactic background is of cosmological interest as a cumulative constraint on stellar nucleosynthesis. Furthermore, it is related to the far-UV background, knowledge of which is crucial to understanding the ionization state of the intergalactic medium. From galaxy number counts in deep images, numerical estimates of the contribution of resolved and unresolved normal galaxies to this background have been made \citep{bernstein02b}. Our present work makes it possible to calibrate a relation between cluster mass and cumulative $U$-band luminosity. By coupling this with a cluster mass function \citep{jenkins}, we can make an independent estimate for the contribution of nearby clusters (with a mass range of $10^{14}$ to $10^{15}$ $M_{\odot}$) to the extragalactic $U$-band background light. Furthermore, assuming that there is no break in the mass-luminosity relation, we can extrapolate it to roughly estimate the contribution from lower-mass, group-like systems ($10^{12}$ to $10^{14}$ $M_{\odot}$).

Determinations of the $U$-band GLF in clusters at low redshift have so far been complicated by the generally low efficiency of detectors in the $U$-band and by the challenges of surveying both wide and deep. \citet{beijersbergen} give a deep determination in the Coma cluster, but use statistical background subtraction rather than a spectroscopically selected sample to account for contamination of the sample by fore- and background galaxies. As \citet{valotto} have demonstrated, this technique can be subject to large systematic errors if the background is inhomogeneous. Spectroscopic samples allow for superior background subtraction.

The recent availability of blue-sensitive wide-field detectors now makes measurement of $U$-band GLFs possible. The combination of a $U$-band photometric survey (McIntosh, in prep.) with a spectroscopic sample of cluster galaxies from the Las Campanas Nearby Cluster Survey \citep{cz2003} enables us to, for the first time, present $U$-band GLFs of three nearby clusters from a spectroscopically selected galaxy sample. The availability of $R$-band photometric data for the same set of galaxies allows us to make self-consistent comparisons between $R$-band and $U$-band GLFs from the same sample.

Our procedure of combining additional photometric data in the $U$-band with an existing survey whose completeness is known in the $R$-band requires us to deal with at least three quantities in the calculation of the GLF: the apparent $U$-band magnitude ($m_{U}$), the $R$-band magnitude ($m_{R}$), and the $R$-band surface brightness ($\mu_{R}$). We have therefore developed a new algorithm for the calculation of the GLF that is particularly suited for such multi-variate analyses. This new algorithm is a variant of maximum likelihood estimators and retains their advantage of being unbiased by density inhomogeneities due to large scale structure. In addition, it offers the benefit that no analytical or binned form for the galaxy parent distribution needs to be assumed {\it a priori}, a great advantage in multi-variate problems, where analytical forms often do not exist and binning the distribution in a multi-dimensional space is inefficient.

We discuss our cluster sample and the spectroscopic and photometric surveys in \S 2. In \S 3 and App. \ref{appdml}, we introduce our new GLF algorithm, the Discrete Maximum Likelihood method, and we discuss the completeness of each survey, as well as systematic corrections to account for biases related to color terms in the sampling fraction in App. \ref{appsf}. \S \ref{secresults} gives our results and discussion. We present the GLFs for the three individual clusters in \S \ref{cluglfs}. We determine the composite GLF for all galaxies as well as for emission line (star forming and active) and non-emission line (quiescent) galaxies in \S \ref{compglfs}. We then compare the $U$- and $R$-band GLFs, which are calculated from the same sample and from the same processing pipeline (\S \ref{compur}). In \S \ref{ubandbackgr}, we examine the contribution of clusters to the metagalactic $U$-band background. In Appendices \ref{apptotallum} and \ref{applargeradii}, we discuss the effects of the spatial and magnitude limits of our survey on our ability to sample the total $U$-band flux from clusters. Our conclusions are presented in \S \ref{conclusions}.

\section{THE DATA}

\subsection{The Cluster Sample}

\newcommand{\degree}{\ensuremath{{}^{\circ}\,}}

Our sample consists of three clusters (Abell 496, Abell 754, Abell 85) from the original spectroscopic and $R$-band imaging sample of \citet{cz2003}. The Christlein \& Zabludoff clusters were selected based on 1) their visibility from Las Campanas, 2) the availability of some prior spectroscopic and X-ray data in the literature, 3) their redshifts, which allowed us to sample the cluster to at least one virial radius with the {1.5\degree$\times$1.5\degree} field of the fiber spectrograph field, and 4) their range of velocity dispersions, which suggest a wide range of virial masses. The properties of these clusters (for $H_{0}=71$ km s$^{-1}$ Mpc$^{-1}$, $\Omega_{m}=0.27$ and $\Omega_{\Lambda}=0.73$, as applied throughout this paper) are given in Table 1. We refer readers to \citet{cz2003} for details on the spectroscopic survey and data reduction.

\begin{deluxetable}{lrrrrcrr}
\tabletypesize{\scriptsize}
\tablecaption{The Cluster Sample}
\tablewidth{0pt}
\tablehead{
\colhead{Cluster}&\colhead{N}&\colhead{$\bar{cz}$ [km/s]}&\colhead{$\Delta m$ [mag]}&\colhead{$cz$ range [km/s]}&\colhead{$\sigma$ [km/s]}&\colhead{center (J2000.)}&\colhead{area [arcmin]}
}
\startdata
A496 &241&$9910\pm48$&35.78&7731  - 11728&$728\pm36$&$04:33:37.84,-13:15:44.5$&$59.9\times58.7$\\
A754 &415&$16369\pm47$&36.90&13362 - 18942&$953\pm40$&$09:08:32.00,-09:37:00.0$&$60.1\times58.5$\\
A85  &280&$16607\pm60$&36.94&13423 - 19737&$993\pm53$&$00:41:50.46,-09:18:11.6$&$60.2\times58.3$\\
\enddata

Notes: $N$ is the number of sampled galaxies per cluster. $cz$ is the mean redshift, $\Delta m$ the distance modulus. Center coordinates and survey area are for the $U$-band photometric survey, which has a smaller coverage than the $R$-band photometric or the spectroscopic survey.
\end{deluxetable}

\subsection{$R$-band Survey}

We derive our master galaxy catalog from a photometric survey of the clusters in the $R$-band \citep{cz2003}. This catalog is complete within certain magnitude and surface brightness limits, so we use it as the reference to estimate the completeness of the spectroscopic and the $U$-band photometric catalogs as a function of $(m_{R},\mu_{R})$. Details of this survey, the image reduction, photometry, and the construction of the master catalog are in \citet{cz2003}. 

\subsection{$U$-band survey}

\label{ubandsurvey}
The $U$-band cluster galaxy data come from wide-field ({1\degree$\times$1\degree}) imaging of A85, A496, and A754 during a January 2000 run using the NOAO Mosaic Imager on the Kitt Peak National Observatory (KPNO) 0.9-meter Telescope.  Complete details of the sample selection, observations, reductions and photometric calibrations are reported in McIntosh, Rix, Caldwell \& Zabludoff (2004, in prep.). The final sample contains $U$-band and $V$-band magnitudes for a total of 631 spectroscopically confirmed cluster member galaxies, and is comparable in depth and in membership coverage to the most comprehensive study of the Coma cluster by \citet{terlevich01}. In physical units, the $U$-band survey region covers about $1.7\times1.6$ Mpc$^{2}$ in A496, $2.7\times2.6$ Mpc$^{2}$ in A754, and $2.7\times2.6$ Mpc$^{2}$ in A85.

The data are well-flattened and carefully corrected to ensure uniformity of the photometric zero point using a customized reduction pipeline that follows standard image reduction techniques and uses the IRAF\footnote{IRAF is distributed by the National Optical Astronomical Observatories, which are operated by AURA, Inc. under contract to the NSF.} environment.  We perform the basic reduction of the $U$-band data using the IRAF {\sc{mscred}} package.  The images are spatially and spectrally flattened ($\lesssim1\%$ deviations globally) using an optimized night-sky flat-field frame constructed from individual science exposures with all objects masked.  We calibrate the data astrometrically to the USNO-v2.0 system \citep{monet96} and photometrically to \citet{johnson66} $U$ magnitudes on the \citet{landolt92} system.

We perform source detection and extraction using SExtractor \citep{bertin} with the following configuration parameters defining our definition of an imaged source: minimum of 5 detected pixels (DEBLEND\_MINAREA) above a background threshold of $3\sigma_{\rm bkg}$ (DETECT\_THRESH), with overlapping sources deblended if the contrast between flux peaks associated with each object is $\geq0.05$ (DEBLEND\_MINCONT).  We confirm that these parameters provide good source detection and deblending by visually inspecting random regions from each image.  We reject sources flagged (FLAGS$\ge4$) as saturated or otherwise bad, and we exclude all sources within 1 arcmin of image edges.  The empirical magnitude limits where the $U$-band source count distributions flatten and turn over are roughly 20.7 
(A85), 21.0 (A496), and 20.5 (A754) mag.  We separate stellar and 
extended sources by fitting a PSF-convolved bulge$+$disk 
model to the light profiles of detected sources using GIM2D \citep{simard02} following the method described in McIntosh et al. (in prep.).  This method is robust to $V=19$ mag, which corresponds to $\sim20.5$ in $U$ for most cluster 
members.

From 975 redshift coordinates within the coverage of the three cluster $U$ images, we find 727 cluster members (with $cz_i=\left<cz\right>_{\rm cl} \pm 3\sigma_{\rm cl}$) and 248 with recessional velocities outside of these bounds.  We cross-correlate the coordinates of member galaxies from the redshift data with the $U$-band source positions and achieve 631 members from our imaging catalog.  
We define image/redshift matches to be the nearest within $5\arcsec$ and find the mean coordinate separation is $<2\arcsec$.  
Only 2.4\% (12) of the 88 redshift positions without a $U$ image detection are 
brighter than $U=19.5$.  Finally, 112 $U<19.5$ extended
sources have no redshifts due to the fractional incompleteness of the
spectroscopic sample at $R>16$ \citep{cz2003}.
Therefore, our $U$-band galaxy catalog from clusters A85, A496, and A754 is 
{\bf $\sim85\%$} complete down to $U\sim19.5$ mag.

We use the dust maps by \citet{schlegel} for extinction corrections and apply a k-correction of $K_{U}=0.065(z/0.02)$, as appropriate for early-type galaxies \citep{pence76}, to the $U$-band magnitudes. We apply this correction to all galaxies in this sample, because cluster galaxies are predominantly early types, individual morphological classifications are not available, and the k-correction for extreme late types differs from the value for early types by less than 0.1 mag even for the most distant sample cluster.

\section{CALCULATING THE LUMINOSITY FUNCTION}

\subsection{A New Method}

\label{glfmethod}

For reconstructing the luminosity function of our survey, we use a new statistical method, which we refer to as the {\it Discrete Maximum Likelihood} Method. The DML retains the advantages of Maximum Likelihood estimators in being unbiased by density inhomogeneities in a sample and is therefore easily applicable to field as well as cluster samples. In contrast to most maximum likelihood-based LF algorithms \citep{sty,eep,blanton}, which which were developed when luminosity alone was the primary variable of interest in determining galaxy distribution function, the DML algorithm does not assume an ansatz (binned or analytic) for the distribution function {\it a priori}, and is therefore independent of the dimensionality of the parameter space of interest. It is therefore ideally applicable to multivariate distribution functions, i.e., the abundance of galaxies as a simultaneous function of luminosity in different filter bands, surface brightness, environment, star formation, morphology, etc. The derivation of the Discrete Maximum Likelihood Method is discussed in detail in Appendix \ref{appdml}.

\subsection{The Sampling Fraction}

\label{sampfrac}

The reconstruction of the GLF, as described in App. \ref{appdml}, requires knowledge of the sample completeness, $f(\vec{x_{h}} \mid F_{i})$. $f$ is the probability that we have both $U$-band photometry and spectroscopic information for a given object with certain physical properties $\vec{x_{n}}$ (e.g., absolute magnitude and surface brightness) if it is in a particular field $F_{i}$ (characterized by redshift, Galactic foreground extinction, position on the sky). Due to the design of our survey, the sampling fraction is not known analytically, but has to be reconstructed from the data. Our $R$-band detection catalog serves as the reference against which we calculate the completeness of the spectroscopic and $U$-band photometric samples. Appendix \ref{appsf} describes the determination of the sampling fraction, as well as systematic corrections that are necessary to account for color selection effects.

\subsection{Consistency checks}

\begin{figure}
\plotone{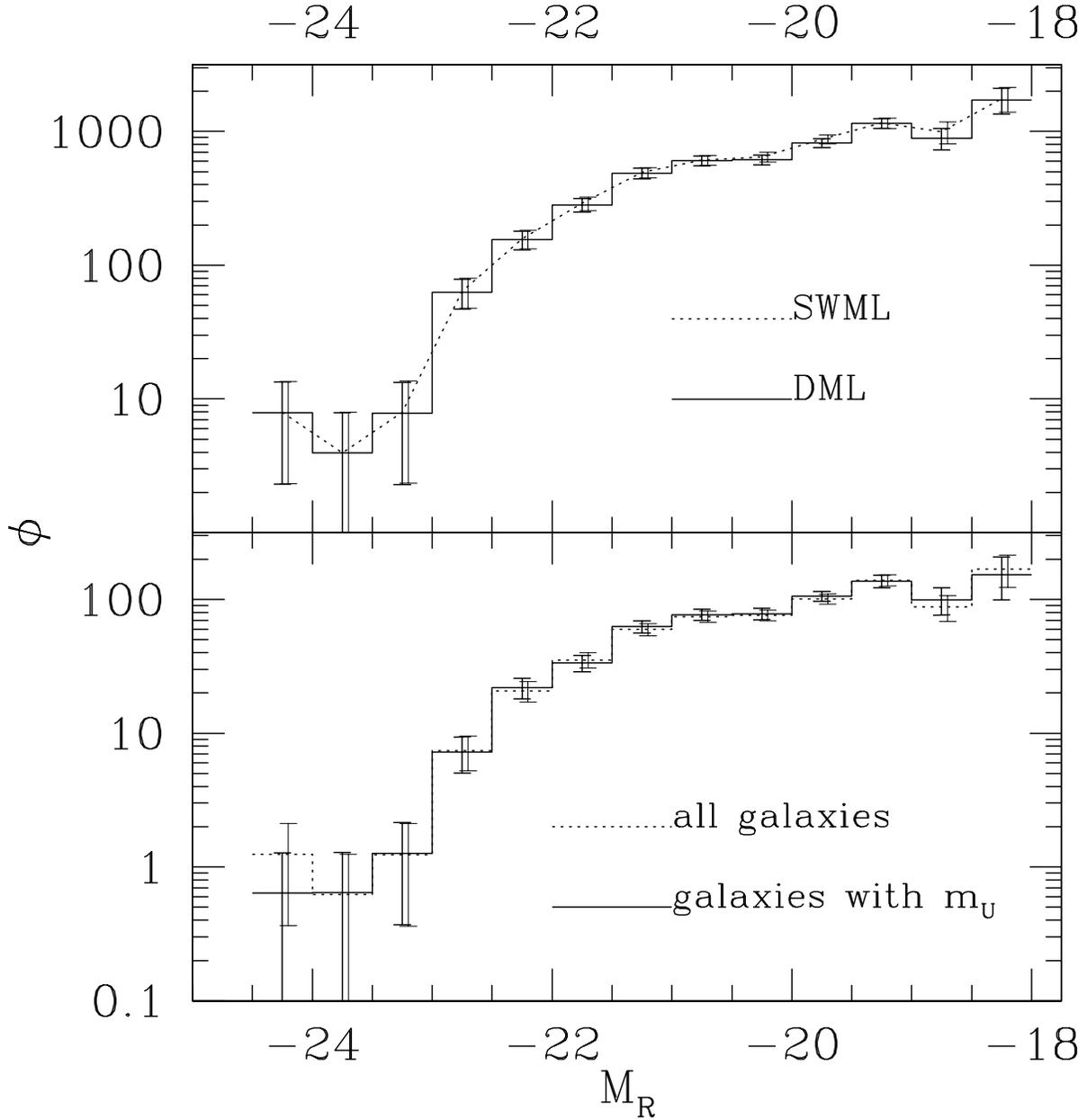}
\caption{Upper pair of curves: Comparison between composite $R$-band cluster GLFs using the Stepwise Maximum Likelihood method (dashed line; Efstathiou, Ellis \& Peterson 1988) and our Discrete Maximum Likelihood method (solid line), determined over the area of the $R$-band photometric sample. Normalization is arbitrary. Lower pair of curves: Comparison between GLFs from all galaxies within $U$-band survey area (dashed line) and from those galaxies with $U$-band photometry only, after completeness correction (solid line). Normalization is galaxies / magnitude / cluster.} 
\label{comp1}
\end{figure}

 The introduction of a new method for the calculation of the luminosity function requires us to demonstrate its consistency with established algorithms. We confirm this by comparing the composite $R$-band GLF obtained from the SWML method (as used in \citet{cz2003}) with our new algorithm. Fig. \ref{comp1} shows these two GLFs (upper curves, unnormalized), determined for the three clusters in our survey. The GLFs are statistically indistinguishable. The results of both methods are equivalent for calculating distribution functions in low-dimensional parameter spaces, but the DML offers greater efficiency and convenience in treating multi-dimensional problems.

 We also test whether our sampling fraction model accounts correctly for the incompleteness of the $U$-band photometric sample. This is particularly important because we model the completeness of the $U$-band catalog as a function of $(m_{R},\mu_{R})$, while the strongest dependence should be on $(m_{U},\mu_{U})$. We calculate the composite $R$-band GLF from all galaxies in the $R$-band master catalog that lie within the spatial boundaries of the $U$-band survey, regardless of whether we know their $U$-band magnitude. We then calculate an $R$-band GLF from galaxies for which we have $U$-band photometry, and apply the appropriate completeness corrections. The agreement between the two GLFs is again excellent, indicating that the incompleteness of the $U$-band catalog does not introduce systematic biases. We emphasize that this $R$-band GLF is plotted using the same weighting and normalization factors as the $U$-band GLF that we present below.

\subsection{Coverage of $(M_{R},M_{U},\mu_{R})$ Parameter Space}

\label{coverage}

Our survey has magnitude limits in two different filter bands, $R$ and $U$. At each $M_{U}$, the $R$ band magnitude limit introduces an incompleteness in the calculated $U$-band GLF, because, for any given $M_{U}$, some fraction of galaxies are too faint in $M_{R}$ to have been sampled.  To avoid this problem, we calculate $U$-band GLFs only down to a limit at which the fraction of lost galaxies is likely to be small.

\begin{figure}
\plotone{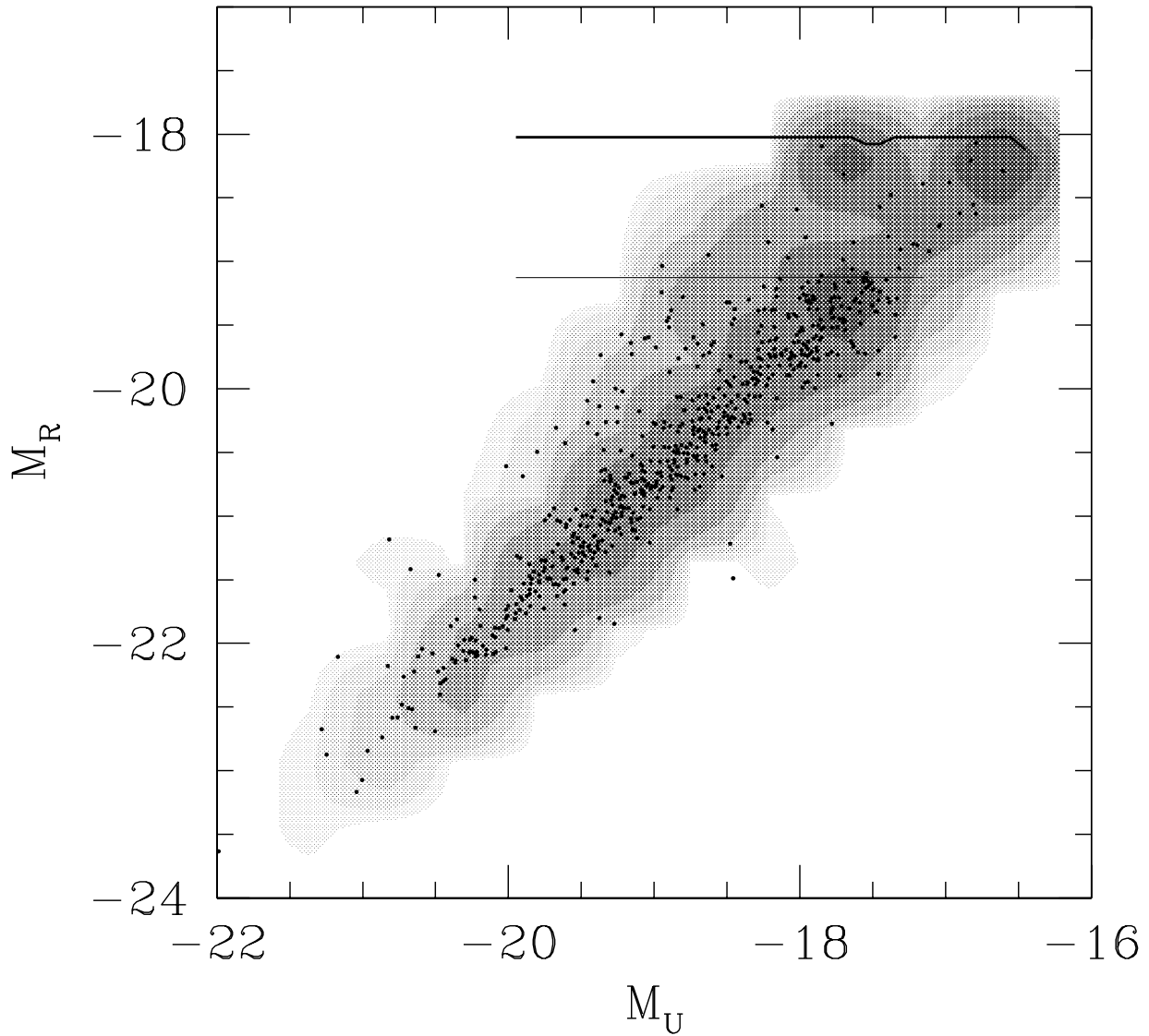}
\caption{Galaxy distribution in the $(M_{U},M_{R})$ plane in greyscale as calculated using the Discrete Maximum Likelihood method. Points denote cluster members with both redshifts and $U$-band photometry available. The bold horizontal line at $M_{R}\approx-18.0$ gives the effective absolute magnitude limit in $R$, i.e., the $M_{R}$ at which the sampling fraction is zero for a galaxy of a given $M_{U}$ and typical $\mu_{R}$ at that $M_{U}$. Galaxies fainter than $M_{U}\approx-17.5$ are too faint in $M_{R}$ to have been sampled, thus motivating our absolute magnitude limit of $M_{U}=-17.5$. The more distant clusters are only sampled to $M_{R}\approx-19.0$ (thin horizontal line), motivating a limit of $M_{U}=-18.5$.}
\label{lfur}
\end{figure}

We use Fig. \ref{lfur} to identify this completeness limit for our $U$-band GLF. The figure shows the completeness-corrected galaxy distribution (greyscale) in the $(M_{U},M_{R})$ plane. Dots represent the individual galaxies in the sample. The bold solid line shows the faintest $M_{R}$ to which a galaxy with a given $M_{U}$ and any $\mu_{R}$ typical for galaxies at that $M_{U}$ could have been sampled. The effective absolute magnitude limit at the faint end is $M_{R}\approx-17.3$. Fig. \ref{lfur} shows that, for $M_{U}>-18$, the galaxy distribution has not been sampled completely; in any $M_{U}$ bin substantially fainter than that limit, a substantial fraction of galaxies are likely to be missing due to the $M_{R}$ limit. We adopt $M_{U}=-17.5$ as the absolute magnitude limit for our analysis of the $U$-band GLF; at this limit, most of the galaxy distribution function has been sampled, and only a small fraction of galaxies ($\sim5\%$, if the galaxy distribution in $M_{R}$ for a given $M_{U}$ is approximately Gaussian) in the faint-$M_{R}$ tail are likely to be lost. The two more distant clusters, A754 and A85, have only been sampled to $M_{R}=-19.0$. Therefore, we consider the individual $U$-band GLFs for these clusters to be reliable down to $M_{U}=-18.5$.

The projection of the distribution onto the $(M_{U},M_{R})$ plane is noticeably broader at fainter magnitudes than at the bright end, a result of the greater fraction of blue, emission line galaxies.

\section{RESULTS AND DISCUSSION}

\label{secresults}

\subsection{$U$-band GLFs for Individual Clusters}

\begin{figure}
\plotone{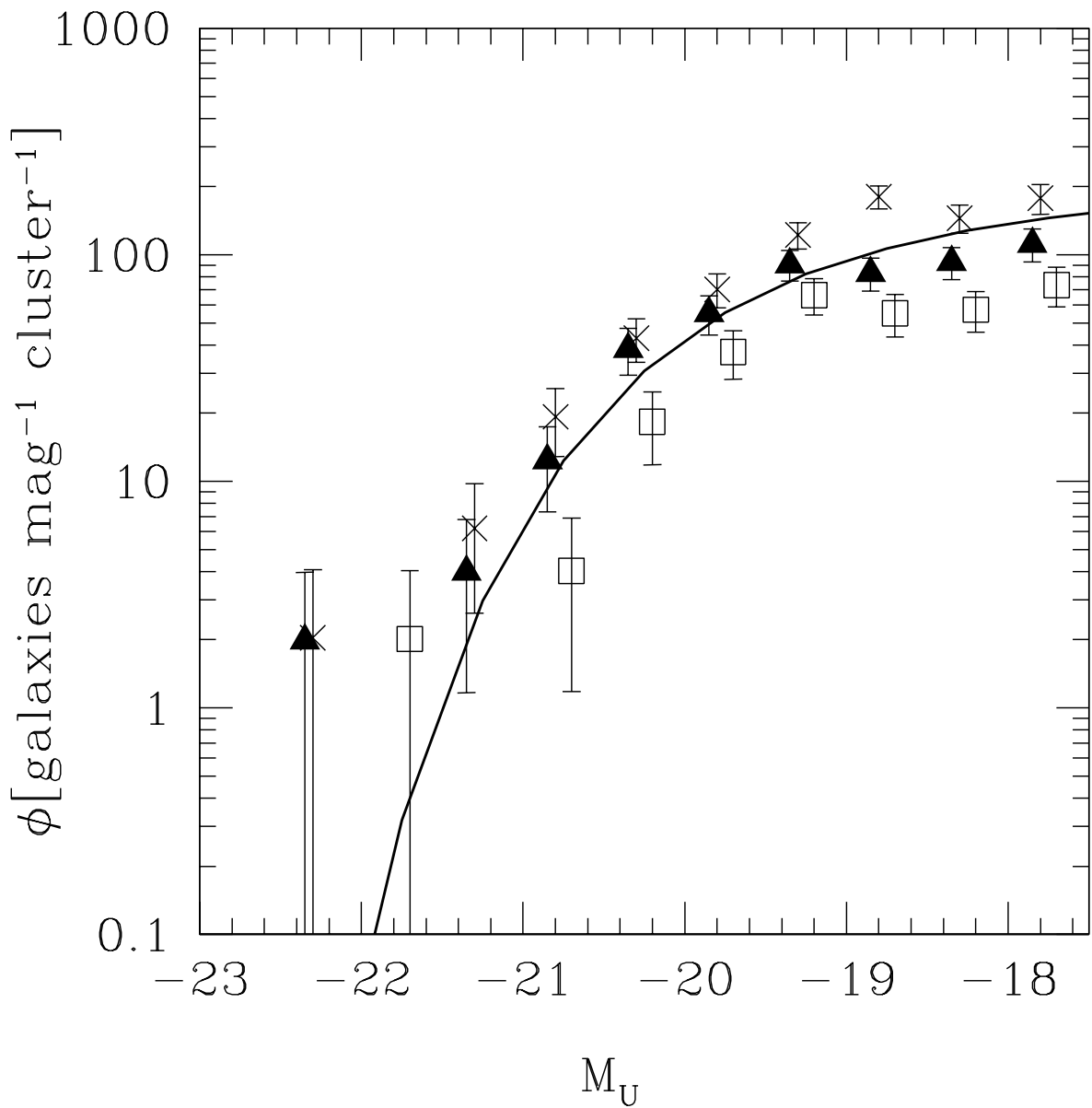}
\caption{$U$-band GLFs for individual clusters. Normalization of all GLFs is preserved. We show the GLFs for A496 (open squares), A754 (crosses), and A85 (triangles). Solid line shows best fit Schechter function to composite GLF.}
\label{lfucls}
\end{figure}

 Fig. \ref{lfucls} shows the individual GLFs for the three clusters (calculated by the method in \S \ref{glfmethod}) superimposed on the composite GLF, which we calculate in \S \ref{compglfs}. Table \ref{tablfucls} gives the numerical values for the cluster GLFs down to the effective sample limits. 

\begin{deluxetable}{lrrr}
\tabletypesize{\scriptsize}
\tablecaption{$U$-band GLFs. Corrected number of galaxies per mag and cluster. Number of sampled galaxies per bin in parentheses.}
\tablewidth{0pt}
\tablehead{
\colhead{$M_{U}$}&\colhead{A496}&\colhead{A754}&\colhead{A85}
}
\startdata
 -22.25&  ...&   2.04(1)&   1.98(1)\\
 -21.75&   2.02(1)&  ...&  ...\\
 -21.25&  ...&   6.21(3)&   3.98(2)\\
 -20.75&   4.03(2)&  19.28(9)&  12.38(6)\\
 -20.25&  18.31(8)&  42.88(21)&  38.42(18)\\
 -19.75&  37.28(17)&  70.49(34)&  55.15(26)\\
 -19.25&  66.42(30)& 122.43(56)&  90.64(42)\\
 -18.75&  55.13(22)& 180.50(73)&  82.94(36)\\
 -18.25&  57.23(24)& 145.09(50)&  92.78(38)\\
 -17.75&  73.61(25)& 177.61(44)& 111.43(36)\\
 \enddata
\label{tablfucls}
\end{deluxetable}

The shapes of the individual cluster GLFs are consistent with each other within $1\sigma$ under a $\chi^{2}$ test if the normalization is adjusted to optimize the agreement. However, the cluster GLFs are significantly offset in normalization. By drawing 1000 Monte Carlo samples of galaxies from the three cluster GLFs, we find the expected number of galaxies at $M_{U}<-18.5$: $91.8\pm11.4$ in A496, $222.0\pm17.4$ in A754, and $143.1\pm13.8$ in A85 over their respective survey regions. If we truncate the survey regions to comparable physical radii (taking into account both the angular diameter distance and the fact that characteristic length scales tend to increase linearly with $\sigma$ \citep{girardi}), the numbers are $91.8\pm11.4$ in A496, $189.2\pm15.8$ in A754, and $120.9\pm12.9$ in A85. The differences between A754 and the other two clusters are significant and most likely reflect the higher mass of A754.

\label{cluglfs}

In the next section, we will form a composite GLF from the three individual clusters. \citet{depropris} find evidence that the shape of the GLF in clusters varies as a function of distance from the cluster center, raising the question whether it is legitimate to form a composite GLF from three clusters with inhomogeneous sampling radii. In our sample, we find no statistically significant differences in the shapes of the GLFs of A754 and A85 between the truncated and untruncated samples, which legitimizes our approach to forming the composite GLF.

\subsection{Composite $U$-band GLFs}

\label{compglfs}
\begin{figure}

\plotone{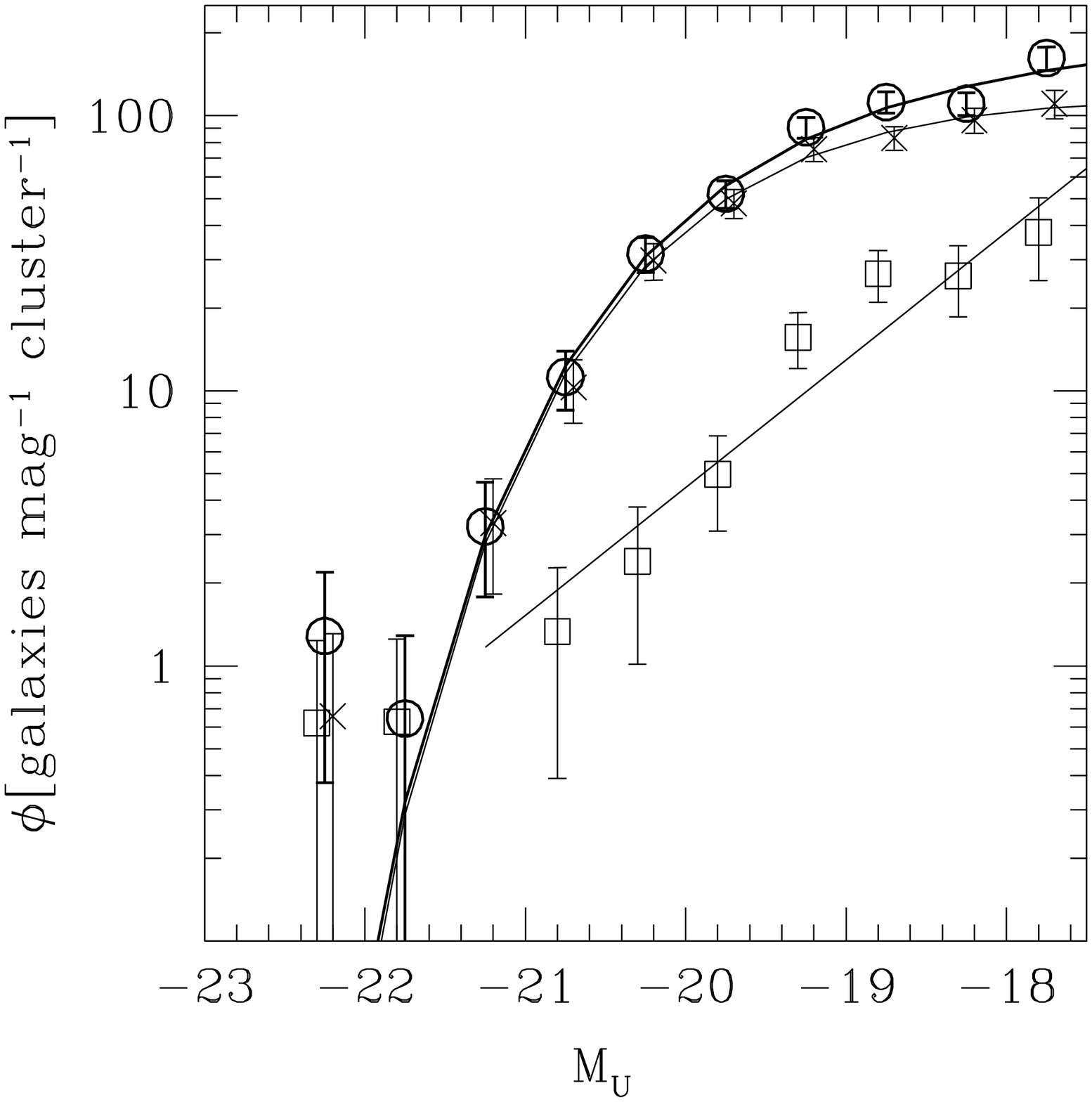}
\caption{$U$-band GLFs for all galaxies (circles), non-emission line galaxies (crosses) and emission line galaxies (open squares). Data points for emission and non-emission line galaxies are displaced by $\pm$0.05 mag for display purposes only. Schechter functions for overall (bold) and non-emission line galaxies, as well as the power law fit for emission line galaxies, are also shown.}
\label{lfucomp}
\end{figure}

We apply the DML algorithm to the complete sample of three clusters and then bin the recovered galaxy distribution function over $M_{U}$. We repeat this process for subsamples of emission line (EL) and non-emission line (NEL) galaxies separately. EL galaxies are defined as having an equivalent width of the [OII] $\lambda 3727$ doublet of {5\AA} or more and are thus star-forming or active galaxies. NEL galaxies have an equivalent width of less than {5\AA} and are hereafter classed as quiescent galaxies. Fig. \ref{lfucomp} shows all three composite $U$-band GLFs. We have applied corrections for B-R color terms, as described in App. \ref{appsf}, in all cases. $1\sigma$ error bars are based on Poisson errors. Table \ref{tabcompglfs} gives the numerical values for the three GLFs, including the numbers of galaxies in each bin. All GLFs are normalized as described in App. \ref{appdml} to represent the average number of galaxies per magnitude and cluster within the region of the $U$-band photometric survey. 

\begin{deluxetable}{lrrr}
\tabletypesize{\scriptsize}
\tablecaption{Composite $U$-band GLFs. Corrected number of galaxies per mag and cluster. Number of sampled galaxies per bin in parentheses.}
\tablewidth{0pt}
\tablehead{
\colhead{$M_{U}$}&\colhead{all}&\colhead{EL}&\colhead{NEL}
}
\startdata
 -22.25&   1.29(2)&   0.62(1)&   0.66(1)\\
 -21.75&   0.64(1)&   0.63(1)&  ...\\
 -21.25&   3.22(5)&  ...&   3.31(5)\\
 -20.75&  11.22(17)&   1.33(2)&  10.29(15)\\
 -20.25&  31.53(47)&   2.40(3)&  29.78(43)\\
 -19.75&  52.03(77)&   4.98(7)&  48.11(68)\\
 -19.25&  90.73(128)&  15.65(19)&  75.71(104)\\
 -18.75& 112.20(131)&  26.65(22)&  83.11(100)\\
 -18.25& 110.55(112)&  26.15(12)&  96.40(89)\\
 -17.75& 161.86(105)&  37.70(9)& 110.45(70)\\
 \enddata
\label{tabcompglfs}
\end{deluxetable}

We have fitted Schechter functions \citep{schechter1976} to these GLFs. The functional form is
\begin{equation}
\phi(M)=\phi^{*}(10^{0.4(M^{*}-M)})^{1+\alpha} exp(-10^{0.4(M^{*}-M)}) ,
\end{equation}

where the nomenclature $\phi$ is used to distinguish this function, which is only a function of absolute magnitude, from the $\varphi$ in Eqn. \ref{dmlansatz}, which is a function of an argument of arbitrary dimensionality.

 The faint end slope is not well-constrained in this sample because the GLFs are only sampled to a fairly bright magnitude limit. The best fit Schechter parameters are $M^{*}_{U}=-19.82^{+0.27}_{-0.27}$, $\alpha_{U}=-1.09^{+0.18}_{-0.18}$, and $\phi^{*}_{U}=142\pm5$ for the overall GLF. Errors in $\phi^{*}_{U}$ are for the best fit $\alpha$ and $M^{*}$ held fixed. Corrections for the B-R color terms in the sampling fraction, as described in App. \ref{appsf}, are $<<1\%$ in $\phi$ and thus negligible for the overall GLF.

The NEL GLF is also described well by a Schechter function. The best fit parameters are $M^{*}_{U}=-19.77^{+0.28}_{-0.30}$, $\alpha_{U}=-0.97^{+0.22}_{-0.18}$ and $\phi^{*}_{U}=133\pm6$ for the NEL GLF. Corrections for the B-R color terms are more important here, because the color difference between EL and NEL galaxies is larger than between field and cluster galaxies. Without our color corrections, the faint end slope would be shallower by $\Delta\alpha\approx0.11$. 

For the EL galaxies, both a Schechter function and a power law provide acceptable fits. To reduce the number of free parameters, and because the Schechter parameters are very weakly constrained by the EL GLF, we decide to fit a power law of the functional form 
\begin{equation}
\phi(M)=\phi^{*} (10^{0.4(-(M+20))})^{1+\alpha}
\end{equation}
The best fit parameters are $\alpha_{U}=-2.16^{+0.16}_{-0.19}$ and $\phi^{*}=4.45^{+0.52}_{-0.52}$. Again, color corrections are important here: the uncorrected EL GLF is steeper by $\Delta\alpha\approx0.12$. 

\subsection{Comparison to $R$-band GLFs}

\label{compur}

Earlier attempts to determine the GLF in clusters have yielded conflicting results, with reported faint end slopes varying from $\alpha\approx-1.0$ to $\alpha\approx-2.2$. One possible explanation of these discrepancies could lie in the use of different filter bands. Comparisons between two recent, spectroscopically selected studies of the cluster GLF \citep{cz2003,depropris} do not show evidence for systematic variations in the faint end slope $\alpha$ between the $R$-band and the $b_{J}$-band. Our present work provides an even more stringent test for the uniformity of the GLF across different filter bands, because the $U$-band is even more sensitive to star formation and dust than the $b_{J}$ band. If differences in the GLFs exist, they are likely to be revealed in a comparison between $U$ and R.

\begin{figure}
\plotone{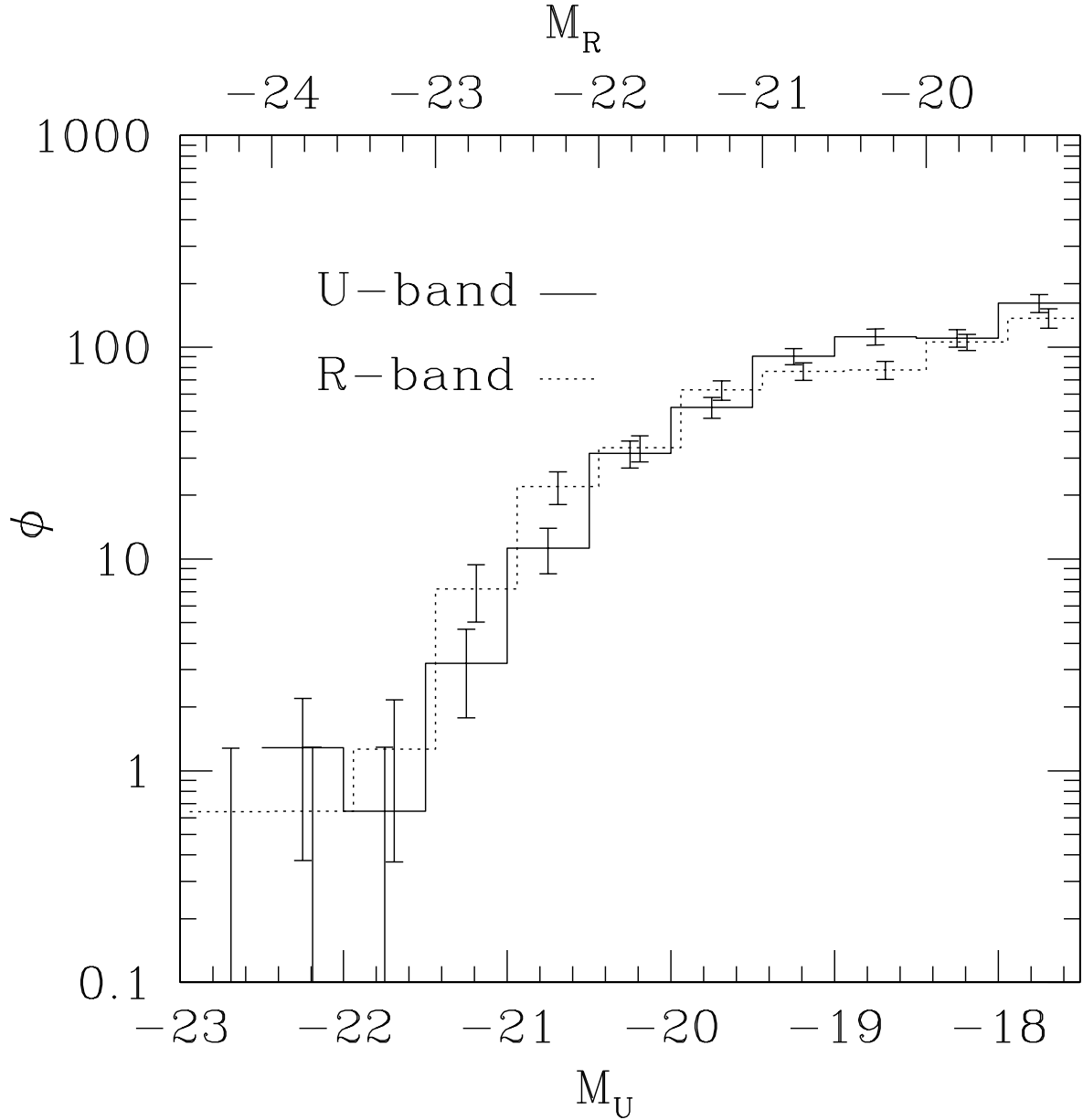}
\caption{Superposition of $U$-band and $R$-band GLFs. The $R$-band GLF (dotted line) has been shifted by $\Delta=U-R=1.56$ mag to maximize the agreement between both GLFs. Detailed analysis shows the $U$-band GLF to be steeper than the $R$-band GLF.}
\label{figlfulfr}
\end{figure}

We first compare the $U$-band and $R$-band GLFs using a $\chi^{2}$ test, allowing for one degree of freedom (the shift in $M$) to minimize $\chi^{2}$. This approach not only tests the GLFs for consistency in shape, but also provides a numerical estimate of the magnitude offsets between the $U$-band and the $R$-band. The magnitude offset that optimizes $\chi^{2}$ is $U-R=1.56\pm0.04$. The maximum probability of consistency is $p=0.10$. Although the $R$- and $U$-band GLFs are formally consistent under this $\chi^{2}$ test, a comparison of the ratio of galaxies with $M^{*}_{U}<M_{U}<-17.5$ to galaxies with $M_{U}\leq M^{*}_{U}$ reveals that the $U$-band GLF is slightly steeper; the ratio is $8.3\pm0.9$ for the $U$-band versus $5.2\pm0.5$ for the $R$-band. Fig. \ref{figlfulfr} shows a superposition of the $U$-band GLF and the shifted $R$-band GLF. 

\begin{figure}
\plotone{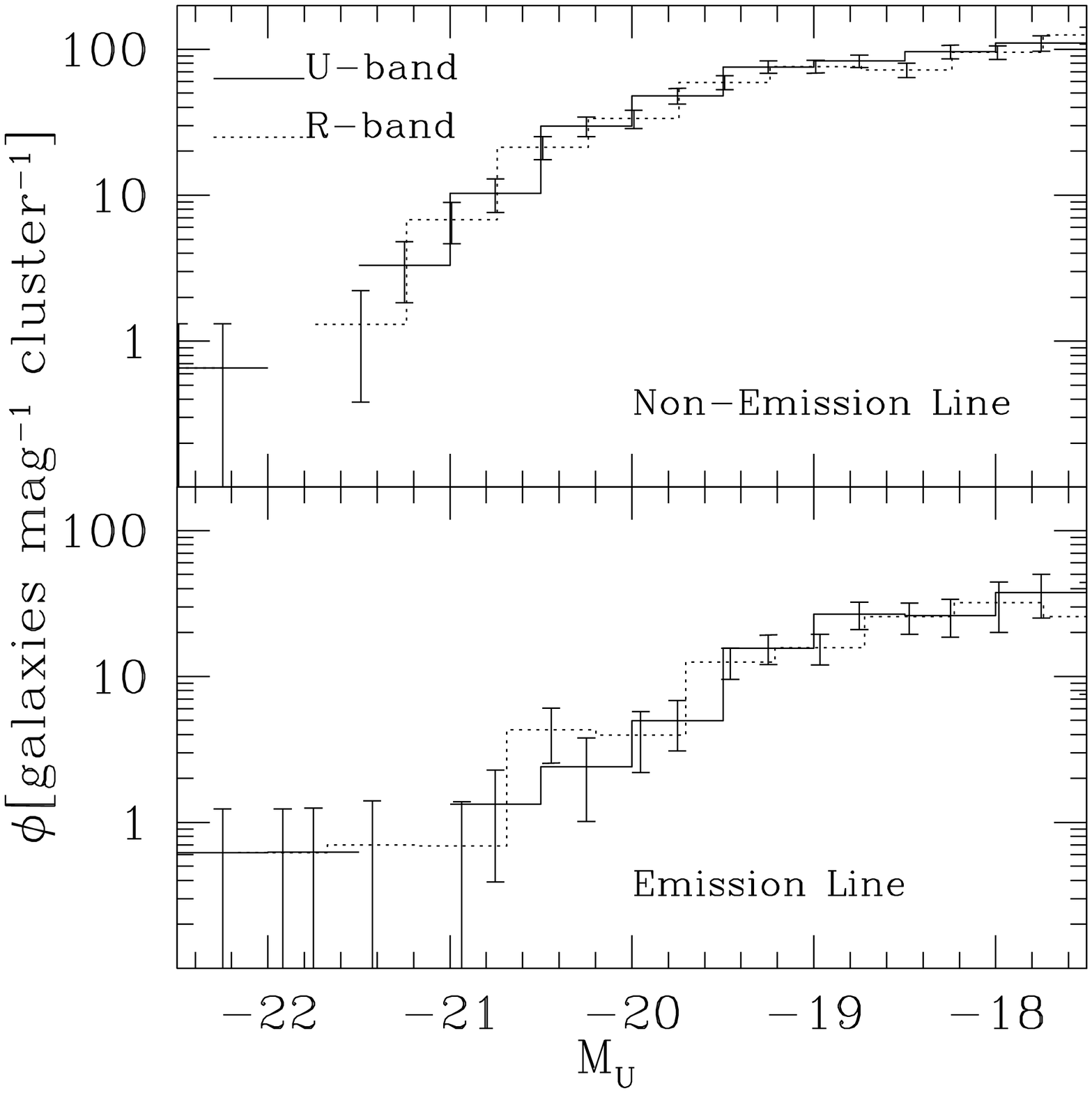}
\caption{Same as Fig. \ref{figlfulfr} for emission line (top) and non-emission line (bottom) galaxies. The $R$-band emission line GLF has been shifted by $\Delta=U-R=0.76$, and the non-emission line GLF by $\Delta=U-R=1.65$. Within each subsample, we find no significant discrepancy in the GLF shapes between the $U$- and $R$-band.}
\label{figlfulfrelnel}
\end{figure}

The difference in the shape of the two GLFs indicates that there is a color gradient, $\frac{\partial(U-R)}{\partial R}$, in the sample: fainter cluster galaxies are bluer. What is the cause of this color gradient? Factors that could influence $(U-R)$ colors are star formation, nuclear activity, dust, and metallicity. Of these, AGN do not appear to play a significant role in our study: the NASA Extragalactic Database lists only one Seyfert galaxy in our sample. We now test whether differences in star formation activity alone could account for the color gradient. For this purpose, we separate our sample into star-forming (EL) and quiescent (NEL) galaxies. If star formation activity is solely responsible for the color gradient, then allowing for star-forming and quiescent galaxies to have different colors should account for the discrepancy in the overall GLF.

 Fig. \ref{figlfulfrelnel} compares the $U$- band and $R$-band GLFs of the EL and NEL subsamples. The optimal magnitude shift that minimizes $\chi^{2}$ for the EL GLFs is $0.76\pm0.30$. The maximum $\chi^{2}$ probability is 0.89, indicating that these GLFs are consistent with each other in shape. For the NEL galaxies, the $\chi^{2}$ comparison yields a magnitude shift of $U-R=1.65\pm0.13$ with a probability of $p=0.97$. The ratio of faint to bright galaxies, determined as above and relative to the same absolute magnitude, $M_{U}^{*}$, is not significantly different between the $U$- and the $R$-band in either the EL or NEL subsample, suggesting that $U-R$ colors are more homogeneous in the subsamples than in the overall sample.

If we shift each of the $R$-band EL and NEL GLFs by their respective best $U-R$ magnitude offset and combine them, the resulting GLF has a steeper faint-to-bright ratio than the original $R$-band GLF, indicating that star formation accounts for some of the discrepancy between the $U$- and $R$-band GLFs. However, the shifted $R$-band and the $U$-band GLF are still significantly different with regard to the ratio of faint to bright galaxies. Therefore, our simple bisection of the sample into star-forming and quiescent galaxies by [OII] EW does not completely explain the color gradient. Other effects, such as dust and metallicity gradients with luminosity, are probably responsible for the residual color gradient, or else the above binning is not fine enough to homogenize colors in each subsample. Clarification of this question will have to wait for future investigations with larger surveys.

Color gradients can introduce differences in the faint end slope between different filter bands. Can these differences account for the wide range of $\alpha$ that has been reported in the literature? To answer this question, we simultaneously fit Schechter functions to both the $U$- and $R$-band GLFs to determine the difference $\alpha_{U}-\alpha_{R}$ and its uncertainty interval. If we fix $M^{*}_{U}-M^{*}_{R}$ at its best fit value of $1.93$, we obtain $\alpha_{U}-\alpha_{R}=+0.03^{+0.14}_{-0.11}$. This is consistent with the expectation that color gradients within the galaxy population introduce only a small difference of $\alpha_{U}-\alpha_{R}=-\frac{\partial(U-R)}{\partial R}(\alpha_{R}+1)$. With an estimated $\frac{\partial(U-R)}{\partial R}=-0.08$, based on the $U-R$ color offsets for EL and NEL galaxies above and the EL/NEL ratios from \citet{cz2003}, the expectation value for our sample is $\alpha_{U}-\alpha_{R}=-0.01$. Therefore, the faint end slope $\alpha$ in clusters is only a weak function of the filter band. This is qualitatively consistent with \citet{paolillo}, who found the faint end slope $\alpha$ to be nearly identical in the $i$-, $r$-, and $g$-bands.

\subsection{Contribution from Clusters and Groups to Near-UV Background}

\label{ubandbackgr}

Understanding the contribution from normal galaxies to this background light provides important constraints on the star formation history of the universe and the relative importance of non-stellar sources, such as AGN or Ly-$\alpha$ recombination radiation from intergalactic gas \citep{tyson95}. Observational estimates for the extragalactic background intensity in the near-UV are on the order of $2-4\times10^{-9}$ erg s$^{-1}$ cm$^{-2}$ sr$^{-1}$ \AA$^{-1}$ \citep{bernstein02,henry99}. Past studies \citep{pozzetti,bernstein02b} have estimated the contributions from field galaxies to this background radiation from number counts in the Hubble Deep Field. Our data provide an independent estimate of the contribution of nearby massive clusters, which are different from the field with regard to their constituent galaxy populations and evolutionary history.

What is the contribution of clusters such as those in our sample to the extragalactic background in the $U$-band? The rate of increase of the $U$-band background intensity from a population of sources with differential luminosity $L_{U,\lambda}$ and differential spatial density $dn/dM$ (where $M$ is the mass) is
\begin{equation}
\frac{dI_{\lambda}}{dt}=\frac{c}{4\pi}\int L_{U,\lambda}(M)\frac{dn}{dM} dM .
\label{eqnlumint}
\end{equation}

An estimate of the mass function, $dn/dM(M)$, is given by \citet{jenkins} from numerical simulations. We use their mass function for cosmological parameters of $H_{0}=70$, $\Omega_{m}=0.3$ and $\Omega_{\lambda}=0.7$. 

$L_{U,\lambda}(M)$ can be constrained from our data. Table \ref{tabculums} shows the sampled $U$-band luminosity within the common magnitude limit of $M_{U}=-18.5$ for all galaxies, for EL (star forming) galaxies and for NEL (quiescent) galaxies. All clusters have been sampled to at least this limit. We discuss possible contributions from fainter galaxies in App. \ref{apptotallum}, but find that, unless the GLF shows a strong upturn at the faint end, we have sampled most of the $U$-band luminosity of the clusters. The $U$-band luminosity is not corrected for radial sampling limits either, but represents the total emission from a cluster's galaxies within our survey area, which is approximately one harmonic radius. The table also gives the cluster mass, $M_{200}$, taken from \citet{reiprich} and based on ROSAT and ASCA measurements of the intracluster gas profile. We estimate errors in the sampled $U$-band luminosity with a Monte Carlo algorithm. 

\citet{beijersbergen} have determined a $U$-band GLF for the Coma Cluster from statistical background subtraction. Integrating their GLF to $M_{U}=-19$ gives a cumulative luminosity of $L_{U}\approx369\times 10^{39}$ erg s$^{-1}$ $\lambda^{-1}$. At a mass of $M_{200}\approx 13.6\times 10^{14} M_{\odot}$, this luminosity is lower than suggested by the extrapolation from the three clusters in our sample, but on the same order of magnitude. Given various systematic differences between these two surveys --- most notably, the different survey areas and the use of statistical background subtraction --- we refrain from imposing this data point as an additional constraint on our $L_{U}(M_{200})$ relation.

\begin{figure}
\plotone{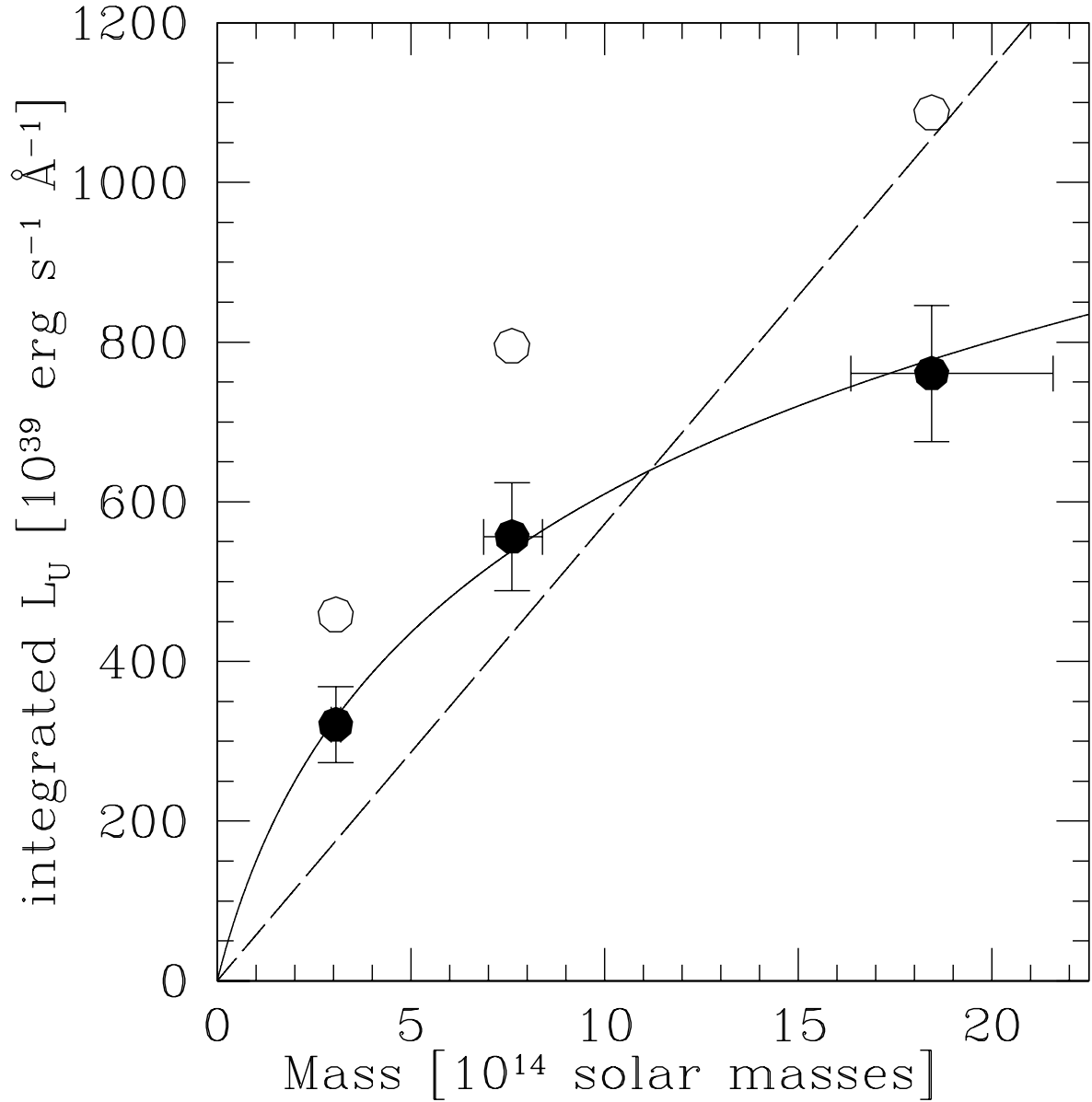}
\caption{Integrated $U$-band luminosity as a function of cluster mass. Filled circles show contribution from the sampled magnitude range ($M_{U}<-18.5$), empty circles show extrapolation to $M_{U}<-10$ using the Schechter fit to the composite GLF. Solid and dashed lines are various analytical fits, which are discussed in \S \ref{ubandbackgr}.}
\label{figcldata}
\end{figure}

\begin{deluxetable}{lrrrrrr}
\tabletypesize{\scriptsize}
\tablecaption{}
\tablewidth{0pt}
\tablehead{
\colhead{cluster}&\colhead{mass [$10^{14}$ $M_{\odot}$]} & \colhead{$L_{U}$ sampled}&\colhead{$L_{U}$ extrapolated}&\colhead{$L_{U}$ extrapolated}
 & \colhead{$L_{U}^{EL}$ sampled} & \colhead{$L_{U}^{NEL}$ sampled} \\
                 &                                       &                          &                              &\colhead{from EL/NEL}                             &                                &                                 \\
}
\startdata
composite &\omit & 533$^{+39}_{-39}$ &  761$^{+  80}_{-70}$ & 1048$^{+ 1714}_{-322}$ &  89$^{+20}_{-20}$ & 447$^{+29}_{-29}$ \\
A496      &3.06$^{+0.13}_{-0.12}$ & 321$^{+48}_{-48}$ &  458$^{+  76}_{-72}$ &  706$^{+ 1453}_{-311}$ &  76$^{+36}_{-36}$ & 241$^{+36}_{-36}$ \\
A754      &18.44$^{+3.13}_{-2.08}$ & 761$^{+85}_{-85}$ & 1087$^{+ 146}_{-135}$ & 1271$^{+ 1408}_{-301}$ &  71$^{+19}_{-19}$ & 682$^{+74}_{-74}$ \\
A85       &7.61$^{+0.79}_{-0.73}$ & 556$^{+68}_{-68}$ &  795$^{+ 113}_{-106}$ & 1156$^{+ 2085}_{-431}$ & 109$^{+44}_{-44}$ & 454$^{+54}_{-54}$ \\
\enddata
\label{tabculums}

Notes: Luminosities in $10^{39}$ erg s$^{-1}$. Errors are based on $1\sigma$ uncertainties in analytical fits. Sampled luminosities are for galaxies with $M_{U}<-18.5$, extrapolated $M_{U}<-10$. Masses are from \citet{reiprich}.
\end{deluxetable}

To derive an estimate for $L_{U,\lambda}$ over a continuous range of cluster masses, we plot the total luminosities from Table \ref{tabculums} in Fig. \ref{figcldata} and fit an analytical expression to the three data points for the uncorrected cumulative luminosity. For similar analyses in the $K$-band, \citet{kochanekwhite01} get $L_{K}\propto M^{1.10\pm0.09}$, consistent with a linear relation. \citet{girardi2000} find a weak, but significant departure from linearity for the $B_{j}$-band. Their estimate is $L_{B_{j}}\propto M^{0.8}-M^{0.9}$. Therefore, previous work suggests both linear and curved relations. To take this uncertainty into account, we attempt to fit both to the data in Fig. \ref{figcldata}, and provide below the range of results from both approaches.

A linear relation does not fit the three clusters in our sample well. The best fit slope in units of $10^{24}$ erg s$^{-1}$ $M_{\odot}^{-1}$ is $57\pm0.7$, with $\chi^{2}=95$.

For the curved relation, in order to avoid a singularity of the mass-to-light ratio at the origin, we do not fit a power law, but opt instead for the functional form
\begin{equation}
L_{U,\lambda}=a_{1}\,lg(a_{2}\,M + 1).
\end{equation}
 This functional form is not theoretically motivated, but it reproduces the data points and provides a plausible interpolation between them, passes through the origin, and is differentiable there. The fit is excellent. With $a_{2}=0.63\times 10^{-14}$ $M_{\odot}^{-1}$, the best-fit value, we find $a_{1}=709^{+88}_{-94}*10^{39}$ erg s$^{-1}$ \AA$^{-1}$.

With these two fits, we obtain luminosity densities from clusters in the mass range from A496 ($M_{200}\approx 3\times10^{14}$ $M_{\odot}$) to A754 ($M_{200}\approx 1.8\times 10^{15}$ $M_{\odot}$) from $1.7$ to $2.8\times 10^{36}$ erg s$^{-1}$ \AA$^{-1}$ Mpc$^{-3}$. The quoted range comprises the difference between the two analytical forms as well as the 1$\sigma$ errors associated with each. This luminosity density contributes to the $U$-band background intensity by $\frac{dI_{\lambda}}{dt}= 4.3-6.4 \times 10^{-12}$ erg s$^{-1}$ cm$^{-2}$ sr$^{-1}$ \AA$^{-1}$ Gyr$^{-1}$, which corresponds to an increase of $0.1-0.3\%$ relative to the current background per Gyr. The higher estimate is associated with the curved functional form.

Estimating the cumulative contribution of clusters to the $U$-band background would require models for the evolution of the cluster mass function, the star formation rate, and the spectral energy distribution of star forming galaxies at high redshifts. A detailed discussion exceeds the scope of this paper, and for this reason, we only give the rate of change in the $U$-band background due to present-day clusters here. However, even if star formation rates in clusters and their progenitors were higher by an order of magnitude in the past (approximately the difference between star formation in present-day field galaxies and the peak of star formation history around $z\approx1$ \citep{madau98,steidel}), the total cumulative contribution of clusters at $z\leq3$ to today's $U$-band background would still be only a few percent. 

Clusters do not contribute much to the mass density of the universe either. Integrating the mass function from \citet{jenkins} shows that clusters in the mass range covered by our sample contribute $\sim2\%$ of the critical density, or $\sim8\%$ of the matter density of the universe. On the other hand, a significant fraction of the $U$-band background may be contributed by galaxies in virialized systems of lower mass, by isolated field galaxies, or by non-stellar sources. Integrating the mass function above shows that groups and clusters with $M_{200}>10^{12}$ $M_{\odot}$ account for a mass density of $\Omega_{m}=0.21$, and additional contributions could come from isolated galaxies and groups with $M_{200}<10^{12}$ $M_{\odot}$.

\begin{figure}
\plotone{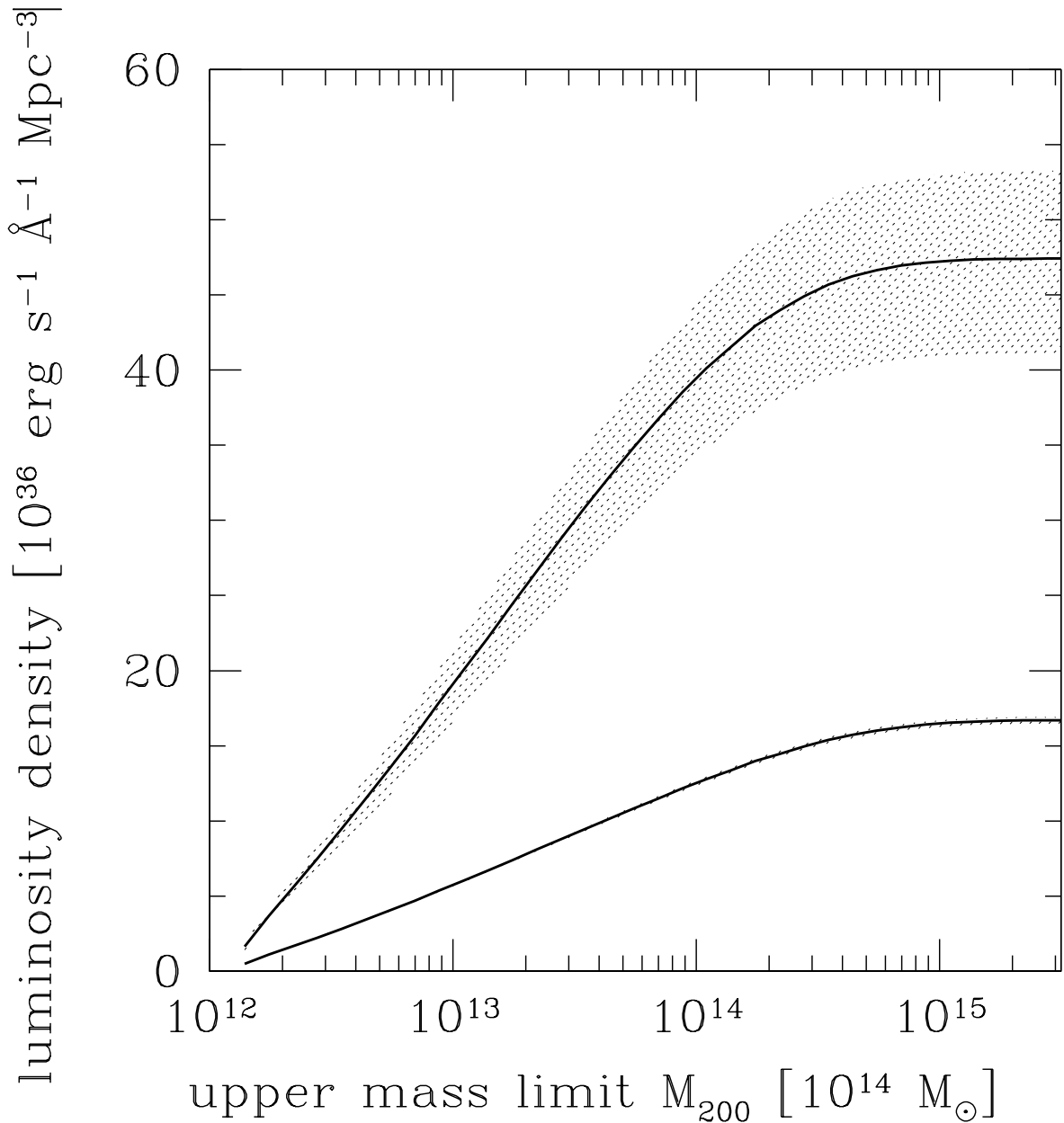}
\caption{$U$-band luminosity density from groups and clusters as a function of the upper mass limit $M_{200}$. Lower mass limit for the integration is $10^{12}$ M$_{\odot}$. Mass function is from \citet{jenkins}.  The upper line corresponds to the extrapolation using a curved functional form, as discussed in the text. The lower line corresponds to a linear relation between $M_{200}$ and $L_{U}$. Uncertainties (shaded regions) are based on the uncertainties in the normalization of $L_{U}-M_{200}$ relation.}
\label{figmintegral}
\end{figure}

To estimate the contribution from groups with $M_{200}>10^{12}$ $M_{\odot}$, we extrapolate the mass-$L_{U}$ relations found above and integrate the luminosity density between $M_{200}=10^{12}$ $M_{\odot}$, the lower bound of the mass of the Local Group, and a variable upper mass cutoff. Fig. \ref{figmintegral} shows the resulting integrated luminosity density as a function of the upper integration limit. The total luminosity density from groups and clusters with $M_{200}\geq 10^{12}$ $M_{\odot}$ approaches $15.0-53.7\times10^{36}$ erg s$^{-1}$ \AA$^{-1}$ Mpc$^{-3}$. Groups of $\sim10^{12}$ to $10^{13}$ M$_{\odot}$ contribute most to this emission. The contribution to the $U$-band background from the entire mass range is $\frac{dI_{\lambda}}{dt}=(42.7-138)\times10^{-12}$ erg s$^{-1}$ cm$^{-2}$ sr$^{-1}$ \AA$^{-1}$ Gyr$^{-1}$, corresponding to an increase of $1-6\%$ of the current $U$-band background over 1 Gyr, depending on the functional form of the fit and the value of the current $U$-band background used.

\section{CONCLUSIONS}

\label{conclusions}

We have calculated galaxy luminosity functions (GLFs) from total $U$-band magnitudes in three nearby clusters. Our analysis is based on a spectroscopic sample, providing cluster membership confirmation for each galaxy and avoiding the need to resort to statistical background subtraction.

We have introduced a new variant of the maximum likelihood method for calculating luminosity functions. Conventional maximum likelihood methods use binned or analytic distributions in just one or two variables (e.g., absolute magnitude and surface brightness) as an {\it ansatz} and require the algorithm to be customized and rerun every time the form of this {\it ansatz} is changed, making it inconvenient to analyze galaxy properties in higher-dimensional parameter spaces (e.g., as a function of absolute magnitude in different filter bands, surface brightness, environment, morphology, star formation, etc., simultaneously). Our {\it Discrete Maximum Likelihood} method, on the other hand, does not assume a specific form or dimensionality for the {\it ansatz} a priori, but assigns a statistical weighting factor to each sampled galaxy. This weighting factor contains all completeness and volume corrections. Therefore, the full range of observables of each galaxy can be retained throughout the calculation and subsequently analyzed without having to customize and rerun the algorithm with a new {\it ansatz} for every new analysis. The DML method is therefore ideal for analyzing modern surveys that gather a large number of photmetric, spectroscopic and morphological parameters about each object. Like other maximum liklihood estimators, the DML has the advantage of being unbiased by density inhomogeneities in the sample.

The results of our GLF analysis are summarized below:

\begin{itemize}

\item The $U$-band GLF in clusters down to $M_{U}<-17.5$ ($\approx M^{*}_{U}+2$) can be described by a Schechter function with $M^{*}_{U}=-19.81\pm0.27$, $\alpha_{U}=-1.09\pm0.18$, and $\phi^{*}_{U}=142\pm5$ galaxies cluster$^{-1}$ mag$^{-1}$. 

\item We compare the $U$-band and $R$-band GLFs and find that, although the difference is too subtle to be reflected in the Schechter parameter $\alpha$, the ratio of faint ($M>M^{*}$) to bright ($M<M^{*}$) galaxies is significantly larger in the $U$-band. This indicates that cluster galaxies are bluer at fainter magnitudes. 

\item For quiescent galaxies, we find best fit Schechter parameters of $M^{*}_{U}=-19.77^{+0.28}_{-0.30}$, $\alpha_{U}=-0.97^{+0.22}_{-0.18}$, and $\phi^{*}_{U}=133\pm6$. The star forming GLF can be fit with a power law with a slope of $\alpha=2.16^{+0.16}_{-0.19}$.

\item If the Schechter fit to the overall GLF can be extrapolated to faint magnitudes ($M_{U}\leq-10$), we have sampled $\sim85\%$ of the total $U$-band light from the clusters within the limit of $M_{U}<-17.5$. Quiescent galaxies dominate the $U$-band flux at $M_{U}<-14$. If there is a faint end upturn in the overall GLF (as suggested by extrapolating the star forming GLF) that continues past $M_{U}=-14$, we cannot rule out that dwarf emission line galaxies dominate the total $U$-band flux. The uncertainty is primarily in the shape of the star forming GLF.

\item Rich clusters in the mass range of our sample ($\sim3\times10^{14}$ to $1.8\times10^{15}$ M$_{\odot}$) account for a $U$-band luminosity density of $(1.71\pm0.02)$ to $(2.46\pm0.33)\times 10^{36}$ erg s$^{-1}$ \AA$^{-1}$ Mpc$^{-3}$. This corresponds to an increase of the current $U$-band background of $0.1-0.3\%$ per Gyr. Additional contributions from galaxies outside our spatial survey limits could increase this estimate by a factor of 2-4. Galaxies beyond our faint magnitude limit could add another $\sim20\%$ if the overall GLF can be extrapolated to $M_{U}\approx-10$. 

\end{itemize}

We thank the anonymous referee for a thorough reading of the text and valuable suggestions. We also thank Dennis Zaritsky for his comments, and Anthony Gonzalez for providing a numerical estimate of the cluster mass function. DC and AIZ acknowledge support from NSF grant \# AST-0206084. This research has made use of the NASA/IPAC Extragalactic Database (NED) which is operated by the Jet Propulsion Laboratory, California Institute of Technology, under contract with the National Aeronautics and Space Administration. 

\appendix

\section{THE DISCRETE MAXIMUM LIKELIHOOD METHOD}
\label{appdml}

Our aim is to recover the galaxy luminosity function, i.e., the parent distribution from which those galaxies with both redshifts and $U$-band photometry in our sample have been drawn. Reconstructing the parent distribution from a set of sampled galaxies requires two corrections. First, a volume correction is needed to account for the fact that bright galaxies are over-represented because they can be observed to greater distances and therefore over a larger volume. This is not a problem in volume-limited surveys (e.g., of only one cluster, where all galaxies are at approximately the same distance), but it is a concern whenever a GLF is derived from a magnitude-limited sample spanning a range in redshift. For example, in our case, a galaxy of a given absolute magnitude may be observable in one cluster, but not in the two other, more distant ones. Second, a completeness correction is necessary because not all galaxies that are photometrically detected in our $R$-band master catalog have been sampled, i.e., have both spectroscopic and $U$-band photometric information. There are various reasons why photometrically detected galaxies are not part of the sample: some faint galaxies were not targeted for spectroscopic observations, a small number were targeted unsuccessfully, and others were not detected on the $U$-band images. The completeness of the sample relative to the $R$-band master catalog varies primarily as a function of apparent magnitude and surface brightness.

GLFs for individual clusters are usually calculated by simply weighting each galaxy by the inverse of its sampling probability, which is either known (in a strictly magnitude-limited survey) or can be recovered empirically (by comparing the number of spectroscopically sampled galaxies of a given magnitude to the total number of photometric detections). Composites are usually calculated by scaling and averaging the GLFs from several individual clusters. However, this method does not make optimal use of the information in a survey. The maximum likelihood solution for the parent distribution function is not only determined by the fact that a galaxy has been observed in one cluster, but also by the question whether or not it would have been observable in any of the other clusters in the sample. 

A method that is typically used for field galaxies, but that could be adapted for a sample of clusters, is the $V/V_{max}$ method, which weights each individual galaxy by the inverse of the volume over which it could have been observed in the survey. This method has two drawbacks: First, it implicitly assumes that the galaxy space density distribution is homogeneous. This is primarily a problem for the field GLF, but if applied to clusters, it also requires introducing an additional scaling to account for the fact that different clusters have different numbers of member galaxies. The second problem is that the method is impractical if the sample is not constrained by a fixed limiting magnitude, but by a fractional completeness that varies with apparent magnitude, which is the case for our cluster data. The definition of the ``volume'' over which a galaxy would have been observable in such a survey is not straightforward.

The most popular algorithms for calculating luminosity functions from a sample spanning a range in redshift are based on maximum likelihood (ML) methods. ML algorithms assume a parent distribution characterized by a limited number of parameters. Taking the selection criteria (sample completeness and magnitude limits) of the survey into account, they then calculate the probability that the observed sample has been drawn from this assumed parent, and iteratively adjust the parameters to maximize this probability. ML methods are preferred because they are not dependent on assumptions about the redshift distribution of galaxies in space and are therefore unbiased by density inhomogeneities due to large scale structure. This advantage applies particularly to determinations of the field GLF, but ML algorithms are applicable to every galaxy sample regardless of the redshift distribution, including clusters. 

 Several variants of the ML method exist that differ in the way they represent the parent distribution. The parametric maximum likelihood method (PML) by \citet{sty} characterizes the parent as a Schechter function \citep{schechter1976} with three free parameters: the shape parameters $M^{*}$ and $\alpha$ and the normalization constant $\phi^{*}$. The disadvantages of this method are that assumptions regarding the shape of the parent are required, and more complex, multi-dimensional analyses are impossible because the Schechter function is a function of only one variable, absolute magnitude. \citet{crossdriver} have expanded the PML to include absolute magnitude and surface brightness. 

\citet{eep}, in the stepwise maximum likelihood (SWML) approach, parametrize the galaxy parent distribution with binned distributions, a method that could, in principle, be expanded to an arbitrary number of dimensions (for an example of an adaptation of the SWML method to two-dimensional distributions, see \citet{cz2003}). However, in a sample of moderate size such as ours, with only a few hundred galaxies, the advantages of binning galaxies are negated by shot noise if the number of bins approaches the number of galaxies in the sample (although the effects of shot noise can be reduced by projecting the distribution back onto a subspace of lesser dimensionality). 

Both the PML and SWML methods suffer from the fact that a parametrized form of the GLF has to be assumed {\it a priori}, and that variables that are not explicitly represented by this form are discarded. They are thus not ideal for multi-dimensional analyses, i.e., for recovering the galaxy distribution function (GDF), which, in analogy to the univariate GLF, describes the abundance of galaxies as a function of multiple variables such as their luminosity in different filter bands, surface brightness, star formation indices, environment, morphology, etc. 

In our case, the fact that we are modeling the sample completeness as a function of $m_{R}$ and $\mu_{R}$, while plotting the luminosity function as a function of $M_{U}$, requires us to keep track of a minimum of three variables $(M_{U},M_{R},\mu_{R})$ and to treat the galaxy distribution function as a function of these (note that in this context $\mu$ is measured in the galaxy rest frame). The PML method is clearly ruled out for our purposes because there is no generally accepted functional form to describe the distribution of galaxies in $(M_{R},m_{U},\mu_{R})$ space.

While the SWML could, in principle, be applied to this problem, it is not very efficient, particularly with regard to future work aimed at expanding our analysis of the galaxy distribution function to higher-dimensional parameter spaces (including such variables as star formation, morphology, and local environment). It either requires modifications to the algorithm to change the ansatz for the galaxy distribution function every time a different cut through the distribution function is to be made, or the distribution has to be binned {\it a priori} in as many dimensions as photometric, spectroscopic and morphological parameters are available for each galaxy. Both approaches are computationally cumbersome, because the {\it ansatz} has to be hardwired into the algorithm and modified every time the dimensionality of the distribution function is changed. Furthermore, there is no advantage to binning a galaxy distribution in so many dimensions that many bins hold only a very small number of galaxies. In such cases, higher-order moments of the galaxy distribution within a single bin can become important, so that the mean properties of the bin do not accurately represent the properties of the individual galaxies contained therein.

We have therefore developed a new variant of the maximum likelihood algorithm that does not require any assumptions about the functional form of the galaxy distribution prior to the calculation, nor does it require binning the galaxies or sacrificing any dimensions of the available parameter space in the sample. 

In our algorithm, the GDF is represented by the sampled galaxies themselves, and not by bins or functional parameters that only model certain aspects of the sampled galaxy population. Our approach can be regarded as an extreme application of the SWML method for infinitesimally small bins in parameter spaces of arbitrary dimensionality. We formally describe the GDF as a sum of weighted delta functions. It is nonzero at all coordinates in parameter space where our sampled galaxies lie. For this reason, we refer to it as the Discrete Maximum Likelihood method (DML) throughout this paper. The {\it ansatz} for the GDF is:
\begin{equation}
\varphi(\vec{x}) = C \sum_{n} \omega_{n} \delta (\vec{x};\vec{x_{n}}) ,
\label{dmlansatz}
\end{equation}
where C is a normalization constant and $\vec{x_{n}}$ is a parameter vector for galaxy $n$ of arbitrary dimensionality. In our case, the parameter space will be $(M_{R},M_{U},\mu_{R})$. The weighting factors $\omega_{n}$ are free parameters to be determined by the DML algorithm. We use the nomenclature $\varphi$ to indicate that this {\it ansatz} can be generalized to arguments $\vec{x}$ of any dimensionality. This {\it ansatz} is similar to that of the {\it C} method \citep{lyndenbell,choloniewski}, but the procedure for solving for the free parameters, $\omega$, is different, with our method retaining the benefits of ML estimators.

The decisive advantage of this method over the SWML method is that, instead of absorbing the individual galaxies into an ansatz for the GDF that represents only a limited number of galaxy properties, it associates a weighting factor with each individual galaxy. The free parameters remain tied to the individual galaxies, rather than to fixed grid points in a pre-selected subspace of parameter space. Therefore, it provides a completeness and volume correction, but still retains the full range of photometric, spectroscopic and morphological properties of the galaxy for a subsequent analysis. Computationally, the algorithm is independent of the dimensionality of the distribution function that is to be analyzed, and only needs to be adjusted to account for the correct dependencies of the sampling fraction. This combines the advantages of the $V/V_{max}$ method with those of the ML methods. The algorithm is also very simple to implement computationally.

The ansatz in Eq. \ref{dmlansatz} is not of course a realistic physical representation of the parent distribution function. After calculating the weighting factors, the distribution can be smoothed or binned, or treated with other multi-variate analysis techniques. In contrast to conventional techniques, these procedures do not have to be hardwired into the algorithm {\it a priori}, but can be applied after the completeness and volume corrections have been carried out, allowing us to perform a multitude of statistical analyses on a data set without having to find a new maximum likelihood solution for each analysis.

The derivation of an algorithm to solve for the free parameters $\omega$ is analogous to the SWML method. We start with the probability that a galaxy $i$ with parameters $\vec{x_{i}}$ would have been observed in the survey:
\begin{equation}
p_{i}= \frac { \varphi(\vec{x_{i}}) f(\vec{x_{i}}\mid F_{i})} { \int \varphi(\vec{x}) f(\vec{x}\mid F_{i}) d\vec{x}} = \frac{ \left( \sum_{g} \omega_{g} \delta(\vec{x_{i}};\vec{x_{g}}) \right) f(\vec{x_{i}}\mid F_{i})}
{\int \left( \sum_{g} \omega_{g} \delta(\vec{x};\vec{x_{g}}) \right) f(\vec{x}\mid F_{i}) d\vec{x}} ,
\end{equation}
where $F_{i}$ describes the field of galaxy $i$ and includes redshift, distance, Galactic extinction, position in the sky, and other information. $f(\vec{x_{i}}\mid F_{j})$ is therefore the probability that a galaxy with properties $\vec{x_{i}}$ would have been included in the sample if it was in the position of galaxy $j$. All indices refer to specific members of the sample of both spectroscopically and $U$-band photometrically sampled galaxies, i.e., $\omega_{h}$ is the weighting factor associated with galaxy $h$. The $\delta$ function now allows us to consider the function only at discrete points and thus solve the integral:
\begin{equation}
p_{i}=\frac{ \left( \sum_{g} \omega_{g} \delta(\vec{x_{i}};\vec{x_{g}}) \right) f(\vec{x_{i}}\mid F_{i})}
{\sum_{g} \omega_{g} f(\vec{x_{g}};F_{i})} .
\end{equation}

We now form the composite probability to have obtained the observed sample from the assumed distribution and take its logarithm:
\begin{equation}
ln {\cal L} = \sum_{i} ln(\sum_{g}\omega_{g}\delta(\vec{x_{i}};\vec{x_{g}})) + 
\sum_{i} ln f(\vec{x_{i}}\mid F_{i}) -
\sum_{i} ln \sum_{g} \omega_{g} f(\vec{x_{g}}\mid F_{i}) .
\end{equation}

Taking the derivative by $\omega_{h}$, we find
\begin{equation}
\frac{\partial ln {\cal L}}{\partial \omega_{h}} =
\sum_{i} \frac{ \delta(\vec{x_{i}};\vec{x_{h}}) }{ \sum_{g}\omega_{g}\delta(\vec{x_{i}};\vec{x_{g}}) } -
\sum_{i} \frac{ f(\vec{x_{h}}\mid \vec{x_{i}}) } { \sum_{g} \omega_{g} f(\vec{x_{g}}\mid F_{i}) }
=
 \frac{ 1 }{ \omega_{h} } -
\sum_{i} \frac{ f(\vec{x_{h}}\mid \vec{x_{i}}) } { \sum_{g} \omega_{g} f(\vec{x_{g}}\mid F_{i}) } .
\end{equation}

Setting this expression to $0$ yields an expression for $\omega_{h}$ that is suitable for an iterative solution:
\begin{equation}
\omega_{h}=\left( \sum_{i} \frac{f(\vec{x_{h}}\mid F_{i})}{\sum_{g}\omega_{g}f(\vec{x_{g}}\mid F_{i})} \right)^{-1} .
\end{equation}

This algorithm converges very quickly. To obtain an absolute normalization factor, we integrate the GDF down to $M_{U}=-17$. We repeat this integration for each galaxy in the sample, applying its visibility conditions $F_{i}$ to predict how many galaxies of a given $M_{U}$ and $\mu_{R}$ would have been sampled, had they been at the position of this galaxy. Thus, we derive a prediction for the total number of galaxies with $M_{U}<-17$ that we would expect to have sampled, and compare it to the actual number to derive a normalization factor:

\begin{equation}
C = N_{sampled} (\sum_{i} N_{cl(i)}^{-1} \sum_{j} \omega_{j} f(\vec{x_{j}} \mid F_{i}))^{-1} ,
\end{equation}
where $N_{cl(i)}$ is the number of galaxies sampled in the cluster to which galaxy $i$ belongs, $N_{sampled}$ is the total number of galaxies in both the spectroscopic and $U$-band sample, and $f(\vec{x_{j}} \mid F_{i})$ is the probability that galaxy $j$ would have been sampled in the survey field and at the redshift and coordinates of galaxy $i$.

Once the weighting factors have been calculated, it is possible to apply techniques of multi-variate analysis and plot different cuts through the GDF over any desired variable, or to restrict the analysis to subsets of galaxies, without having to modify the algorithm or recalculate the weighting factors. In principle, GLFs in the $U$-band and the $R$-band can be plotted from the same output, using the same weighting factors and normalization constants. Furthermore, a sample can be split (e.g., by emission line properties), and the two distributions can be plotted and compared separately; the weighting factors do not have to be recalculated. Therefore, correct relative normalization between the two subsamples is automatically guaranteed.

Care has to be applied in cases where, due to survey design, the sampling fraction model does not fully account for all dependences of the sample completeness. In our study, spectroscopic targets have been selected by approximate $b_{J}$ magnitudes. This introduces a residual dependence of the sampling fraction $f$ on $b_{J}-R$ colors that is not accounted for in our approach, which only models $f$ as a function of $(m_{R},\mu_{R})$. If a subset of galaxies is selected by parameters that are correlated with $b_{J}-R$ colors, such as cluster membership or emission line properties, biases can result, because the mean empirical sampling fraction calculated for galaxies of a given $(m_{R},\mu_{R})$ may be systematically different from the true sample completeness for the selected subset of galaxies. We resolve this problem by applying systematic corrections specific to the individual subsamples (particularly the EL and NEL subsamples) to the sampling fraction, which requires us to recalculate the weighting factors for these subsets. These corrections are discussed in App. \ref{appsf}.

For a similar reason, when calculating the GDF for individual clusters out of a sample of multiple clusters, it is not legitimate to first calculate the weighting factors for the entire sample and then just plot the galaxies associated with one particular cluster, because the weighting factors determined that way would contain volume corrections that are appropriate to a sample spanning a range in redshift, but not to the sample of an individual cluster. For a given absolute magnitude, the sampling fraction is different from cluster to cluster. Therefore, the weighting factors have to be recalculated for each individual cluster in order to analyze individual cluster GDFs. This is not a drawback of the DML method, but common to all ML methods.

When the recovered value of the GDF is exactly zero in any region of parameter space (for example, for very faint $M_{R}$), it is necessary to understand whether this is because the sampling fraction is so low that none of a potentially large population of galaxies have been sampled, or because there are no galaxies with these properties. Again, this is a problem common to all LF algorithms, but a careful treatment is particularly important in the DML, because it represents the GDF solely with sampled galaxies and therefore, by default, does not probe the sampling fraction outside of the sampled regions of parameter space (whereas other methods like the SWML will return an undefined result if the sampling fraction is zero). 

The sampling probability for a galaxy of any given absolute magnitude, surface brightness, etc., can be probed in the DML by introducing mock galaxies. The critical value for the sampling probability can be defined in the following way: Consider the the GDF in two dimensions, $M_{U}$ and $M_{R}$, and define a bin in $M_{U}$. The number of galaxies in the well-sampled parts of the galaxy distribution in this bin, which is represented by the reconstructed GLF, is $N_{0}$. Further, assume that there is a region of parameter space within this $M_{U}$ bin with a low sampling probability. The number of galaxies in this region is N, and the sampling probability for such a galaxy is $f$. The poorly sampled region is a problem only if a) the expected number of sampled galaxies in this region is less than 1, i.e. $N f<1$, and b) the number of galaxies that exist in this region is a substantial fraction $p$ of the total number of galaxies in this $M_{U}$ bin, i.e. $N>p (N+N_{0})> p N_{0}$. This yields the condition for poor sampling, $p N_{0}<f^{-1}$ (note that this expression is independent of the size of the hypothetical poorly sampled region). The reconstructed luminosity function is therefore not affected by a systematic relative error of more than $p$ as long as the sampling fraction $f$ is greater than the critical value of $(p N_{0})^{-1}$ everywhere within this bin. By introducing mock galaxies with the given $M_{U}$, a value of $\mu_{R}$ characteristic for galaxies with this $M_{U}$, and arbitrary $M_{R}$, this requirement can be checked easily and the effective limits of the survey in $M_{R}$ determined. 

Once we have found the effective survey limits (for example, in $M_{R}$), it is necessary to ascertain whether it is likely that substantial fractions of the galaxy distribution lie outside these limits. For example, for any $M_{U}$, there are extreme values of $M_{R}$ for which no galaxies could have been sampled spectroscopically. However, there is a fairly tight correlation between $M_{U}$ and $M_{R}$, and for bright $M_{U}$, it is extremely unlikely that any galaxies reside beyond the faint effective limit in $M_{R}$. In \S \ref{coverage}, we use the procedure outlined above to define the effective absolute magnitude limits in $M_{R}$ and $M_{U}$ for our sample.

\section{DETERMINATION OF THE SAMPLING FRACTION}

\label{appsf}

This section describes the empirical determination of the sampling fraction, $f(\vec{x_{h}} \mid F_{i})$, which is the probability that we have both $U$-band photometry and spectroscopic information for a given object with certain physical properties $\vec{x_{n}}$ (e.g., absolute magnitude and surface brightness) if it is in a particular field $F_{i}$ (characterized by redshift, Galactic foreground extinction, position on the sky). We determine the completeness of the $U$-band and spectroscopic catalogs relative to the $R$-band photometric catalog.  We assume the $R$-band catalog to be complete for $m_{R}<18$ and $\mu_{R}<23.61$ (the approximate equivalent of a per-pixel detection threshold of 1.5$\sigma$ on a typical frame). This assumption is conservative; in fact, we detect objects more than five magnitudes fainter, but do not include them in the analysis. We cannot detect objects at lower surface brightness, but at $m_{R}<18$, the distribution of photometric detections in the $(m_{R},\mu_{R})$ plane is well separated from this limit, indicating that few objects are likely to be lost to the surface brightness limit.

 Following Bayes's theorem \citep{bayes}, we factorize the sampling fraction --- the probability that, for a given object, redshift information ($cz$) and $U$-band photometry ($U$) are available on condition that the object meets our sample selection criteria. These selection criteria, denoted by $sel$, are the criteria that we apply to our galaxy sample to select those galaxies that are to be included in our GLFs. This is primarily cluster membership. For some analyses, we also select galaxies by their emission line properties, and comment on this at the end of this section.

The expression for the conditional sampling probability is then
\begin{equation}
p(cz \land U \mid sel) = p(cz \land U \land sel) / p(sel) .
\label{eqnsampprob}
\end{equation}

We now transform Eqn. \ref{eqnsampprob} into a form that is suitable for numerical evaluation and factorize it into separate terms for the spectroscopic and the $U$-band photometric completeness. There are two straightforward factorizations that allow us to separate terms connected to the spectroscopic and the $U$-band sampling fractions:
\begin{equation}
p(cz \land U \mid sel) = p(U \mid cz \land sel) \times p(cz \mid sel) .
\label{eqnsffactorization}
\end{equation}
or
\begin{equation}
p(cz \land U \mid sel) = p(U) \times p(cz \land sel \mid U) \times p(cz \mid sel) / p(cz \land sel) .
\end{equation}

The separation of terms for the spectroscopic and $U$-band photometric sampling fractions is motivated by the opportunity to apply specific, systematic corrections to either. It is only these corrections that introduce any difference between the two methods, but the differences are statistically insignificant ($<1\%$ even in the faintest magnitude bins). We use the first factorization because of its greater simplicity.

Both the spectroscopic and $U$-band sampling fractions cannot be described analytically in our case. We thus have to determine them empirically by comparing the number counts of all detections in the $R$-band photometric master catalog with those with redshifts and $U$-band photometry. Because the $R$-band photometric catalog is complete, we calculate the sampling fraction as a function of $m_{R}$ and $\mu_{R}$. These two parameters alone do not allow for an unambiguous description of the sampling fraction; color and/or environment also play a role. We discuss below how we address this problem. The bin widths that we use to calculate the sampling fraction are $\Delta m=0.75$ and $\Delta \mu =0.25$. We have experimentally found that these choices of bin widths reproduce stable results for the resulting GLFs. 

$p(cz \mid sel)$ cannot be determined exactly, because information about cluster membership is only available for spectroscopically sampled objects. The assumption $p(cz \mid sel) \approx p(cz)$ is only justified if the sampling fraction is not strongly correlated with the selection criteria for a given $(m_{R},\mu_{R})$.  Unfortunately, the experimental design of our survey --- target selection in $b_{J}$, compilation of a master catalog in $R$ --- violates that assumption, because there are systematic differences in $B-R$ colors between cluster and field galaxies, and therefore systematic differences in the spectroscopic sampling probability for a given $(m_{R},\mu_{R})$. This color selection bias could be remedied if the sampling fraction model explicitly contained color terms, but it is impractical to estimate the sampling fraction as a function of more than two parameters with a sample the size of ours.

Instead, we use an algorithm that systematically compensates for color selection bias and estimates $p(cz\mid sel)$ under the assumption that the targeting probability is uncorrelated with the selection criteria for a given $m_{b_{J}}$. The probability that a galaxy of a given $m_{R}$ obeying the selection criteria has been targeted is 
\begin{eqnarray}
p(target \mid sel)\mid m_{R}=(N_{target,nonsel}^{0}N^{+}+N_{target,sel}^{+}N^{0}+2N^{0}N_{target,nonsel}^{+}+\nonumber\\
2N_{target,sel}^{0}N^{+})+((N_{target,nonsel}^{0}N^{+}+N_{target,sel}^{+}N^{0}+2N^{0}N_{target,nonsel}^{+}+\nonumber\\
2N_{target,sel}^{0}N^{+})^{2}-8N^{0}N^{+}(2N_{target,sel}^{0}N_{target,nonsel}^{+}+N_{target,nonsel}^{0}N_{target,nonsel}^{+}+\nonumber\\
N_{target,sel}^{0}N_{target,sel}^{+}))^{1/2}/4N^{0}N^{+} .
\end{eqnarray}
Here, $N$ denotes the number of galaxies that have been photometrically detected and thus included in the $R$-band master catalog. $N_{target}$ is the number of galaxies targeted for spectroscopic observations. The subscript $sel$ indicates galaxies that are to be included in the GLF (e.g., that are cluster members). The superscript $0$ indicates that these quantities are evaluated at $m_{R}$, and the superscript $+$ means that they are evaluated at $m_{R}+\Delta(B-R)$. $\Delta(B-R)$ is the difference in B-R colors between galaxies that have been selected (e.g., that are cluster members), and those that have not. We roughly estimate the $m_{R}$-dependent differences in $\Delta B-R$ between cluster and field galaxies from the sampled galaxies themselves.

 In all cases, $b_{J}$ is not an accurate photometric magnitude, but rather the approximate magnitude that served as the basis for the target selection for the spectroscopic survey. Furthermore, for this derivation we assume $N_{target,sel}\approx N_{target}N_{cz,sel}/N_{cz}$, i.e., that the spectroscopic success rate for a given $(m_{R},\mu_{R})$ is not strongly correlated with the subsample selection criteria. Any error in this assumption is likely to be small, because the success rate is very high for the magnitude range analyzed. We then modify the sampling fraction $f_{s}(m_{R},\mu_{R})$ by the factor $p(target \mid sel)/p(target) \mid_{m_{R}}$. We generally apply these corrections to our results, except where we explicitly specify otherwise in order to estimate the size of the effect.

Ultimately, we cut our galaxy sample not just by cluster membership, but also by emission line properties. The color selection bias described above applies to this case as well, because emission line (EL) galaxies have systematically different $(B-R)$ colors than non-emission line (NEL) galaxies. Therefore, the probability that a galaxy has been targeted for spectroscopic observations is systematically higher if it is an EL galaxy. We remedy this problem in the same way as described above, by assuming a $\Delta(B-R)$ color difference between EL and NEL galaxies and applying our systematic correction to recover an estimate for $p(cz\mid sel)$.

In principle, a similar correction would be required to account for the fact that the completeness of the $U$-band catalog has a residual dependence on $U-R$. Although Fig. \ref{comp1} shows our compensation for the incompleteness of the $U$-band sample to work very well in reproducing the correct $R$-band GLF, systematic errors arising from this problem are more likely to affect the $U$-band GLF. To quantify the impact that these errors may have on our results, we pursue the following approach:

We model the completeness of the $U$-band photometric sample, as
\begin{eqnarray}
p(U \mid cz \land sel) \equiv f_{U} \equiv f_{U}(m_{R},m_{U},\mu_{R}) \nonumber\\ 
\approx \bar{f_{U}}(m_{R},\mu_{R}) + \frac{\partial f_{U}}{\partial m_{U}}(m_{U}-\bar{m}_{U\mid_{R}}) \nonumber\\
= \bar{f_{U}}(m_{R},\mu_{R}) + \frac{\partial f_{U}}{\partial m_{R}}\frac{d R}{d U}(m_{U}-\bar{m}_{U\mid _{R}}) .
\end{eqnarray}

We determine $\frac{\partial f}{\partial m_{R}}$ from our default sampling fraction model and obtain $\frac{dR}{dU}$ (which is of order unity) directly from the sample. $(m_{U}-\bar{m_{U}}\mid _{R})$ is the difference between the $m_{U}$ for which we determine the sample completeness and the mean $\bar{m_{U}}$ appropriate for the given $m_{R}$. Therefore, this approach uses a first-order approximation to recover the sample completeness for a given $m_{U}$ and $\mu_{R}$ instead of $m_{R}$ and $\mu_{R}$ (thus removing most of the bias, as the dependence on apparent magnitude is stronger than that on surface brightness). 

The resulting $U$-band GLF is consistent with the $U$-band GLF without these corrections ($p_{\chi^{2}}=0.994$). Because the systematic effects are negligible and the above argument is only an approximation to assess the order of magnitude of the effect, we choose not to apply these corrections in our calculations.

\section{HOW MUCH LIGHT FROM FAINTER GALAXIES?}

Our photometric and spectroscopic samples are limited in apparent magnitude and sky coverage. Because of these constraints, we have sampled only a part of the total galaxy population in each of our three clusters. In this and the following section, we will estimate what additional contributions in terms of numbers of galaxies and cumulative luminosity could come from cluster galaxies outside our photometric and spatial survey limits.

 Our composite GLF reaches to $M_{U}=-17.5$. For a GLF with a rising faint end slope, such as the one in our sample, there are many more faint than bright galaxies. On the other hand, most of these are dwarfs that contribute only little to the total $U$-band light from a cluster. How much of the total $U$-band light have we sampled within our survey limit? To quantify this fraction, we fit a functional form to the sampled data and extrapolate it beyond the magnitude limit. As functional forms, we use our Schechter functions for the overall and NEL GLFs, and power laws for the EL GLFs. There are two possible approaches: We can fit a Schechter function to the overall GLF and extrapolate it to a very faint cutoff magnitude, or we can fit functional forms to the EL and NEL GLFs, extrapolate them individually, and add the results.

 These two approaches are generally not equivalent, because the sum of the power law EL and Schechter NEL GLFs is not necessarily a Schechter function. While a single Schechter function, as we have fitted to our overall GLF, assumes that the faint end is approximated by a single power law, the superposition of an EL and NEL GLF --- assuming that there is no break or turnover in either GLF --- usually shows a strong upturn at the faint end, just beyond our sampling limit, due to the increasing dominance of the power law EL GLF over the Schechter NEL GLF with its flat faint end slope. As we do not know whether the overall, EL, or NEL GLFs can be extrapolated all the way to the lower end of the galaxy magnitude range, or whether there is a break or turnover at a magnitude fainter than our survey limit, it is not clear which of these approaches best represents the real GLF of cluster galaxies.

 Claims of an upturn at the faint end of the cluster GLF have been made in the past on the basis of photometric surveys that employed statistical subtraction rather than spectroscopic membership confirmation for background decontamination (\citet{valotto} and references therein). Recent spectroscopic surveys of the GLF in clusters \citep{cz2003,depropris} have not turned up evidence for such an upturn in the $R-$ and $b_J$-bands, suggesting that previous claims may have been affected by biases inherent in statistical background subtraction \citep{valotto}. Some recent photometric surveys of nearby clusters \citep{trentham02} also indicate that the faint end slope is substantially flatter than our estimate for the EL GLF. Nevertheless, the question is not yet settled, because the spectroscopic surveys are generally shallower than purely photometric ones and the results from photometric surveys are still ambiguous. The uncertainty in the slope of the EL GLF in our present study is too large to predict at which magnitude an upturn might become significant.

Because of this uncertainty, we pursue both extrapolation approaches and compare the results below. To estimate our uncertainties, we perform the extrapolations for a range of Schechter parameters within the 1$\sigma$ error contours for the individual fits to the overall, EL, and NEL GLFs.

\begin{figure}
\plotone{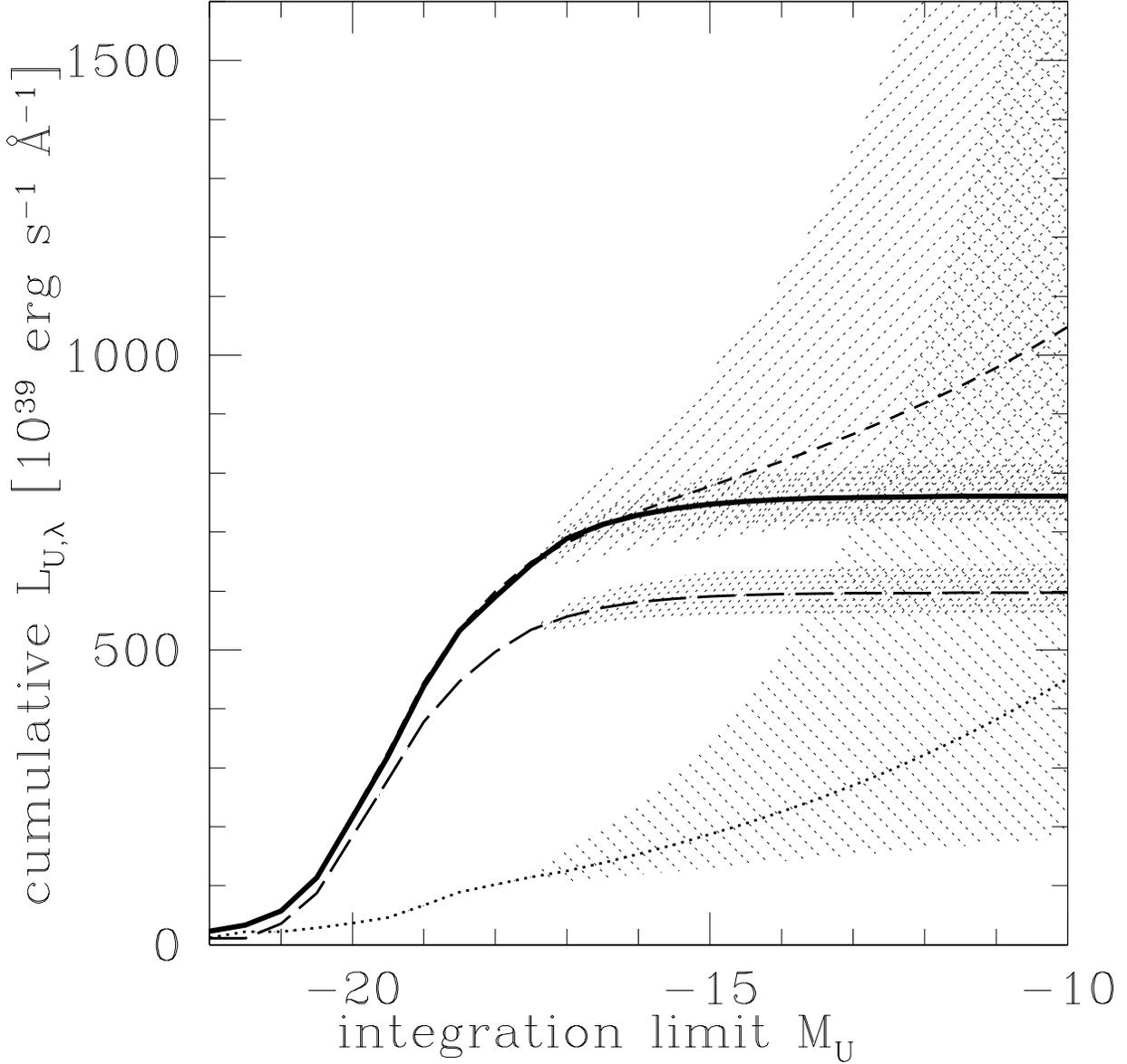}
\caption{Integrated $U$-band luminosity for emission line (EL), non-emission line (NEL) and overall GLFs per cluster as a function of faint integration limit. The solid bold line shows the cumulative $U$-band luminosity of all galaxies, using the Schechter fit for the overall GLF for extrapolation. The long-dashed line shows the same for NEL galaxies. The dotted line shows EL galaxies. The upper, short-dashed line gives the sum of the EL and NEL integrals.}
\label{figculum}
\end{figure}

Fig. \ref{figculum} shows the results for the cumulative $U$-band luminosity per cluster (averaged over the three clusters in our sample) as a function of the limit $M_{U}$ to which we integrate the GLF. We use the zero-point flux for the Johnson $U$-band given by \citet{colina} to convert absolute magnitudes to luminosities. The luminosity of a source with $M_{U}=-19$ is given by $L_{U}^{-19}=2.065\times 10^{39}$ erg s$^{-1}$ \AA$^{-1}$. The faintest known dwarf galaxies with recent star formation in the Local Group have $M_{V}\approx-10$ \citep{pritchet99}. We adopt this value in the $U$-band as the faintest magnitude to which we extrapolate the GLF. 

Judging from the overall GLF, galaxies fainter than our survey limit of $M_{U}=-17.5$ contribute very little to the total cluster $U$-band luminosity. Therefore, we have sampled most of the cluster $U$-band light in our survey. Extrapolating the GLF to $M_{U}=-10$, the fraction of light in the bright end ($M_{U}<-17.5$) is $0.85_{+0.05}^{-0.06}$ (errors based on the $1\sigma$ errors in $\alpha_{U}$ and $M^{*}_{U}$). It is notable that most ($83\%$) of the $U$-band light at $M_{U}<-17.5$ is contributed by giant NEL galaxies, which make up $\sim80\%$ in number. NEL galaxies dominate the $U$-band emission from clusters at least down to $M_{U}=-14$.

However, if we extrapolate the EL and NEL GLFs individually, the sum of the extrapolated GLFs, for $M_{U}>-17$.5, is not as flat as the fit to the overall GLF. The sum shows a strong upturn at fainter magnitudes due to the dominance of the power law EL GLF at faint magnitudes. In this case, populations of EL galaxies much fainter than $M_{U}=-17.5$ contribute substantially to the total $U$-band luminosity of a cluster {\it if there is no break in the EL GLF}. In fact, for a faint end slope $\alpha<-2$, the luminosity integral diverges. If we allow for the existence of a steep faint end upturn that arises from the extrapolation of the EL GLF to fainter magnitudes, our best estimate for the fraction of the light sampled at $M_{U}<-17.5$ is $0.62^{+0.25}_{-0.38}$. In the worst case, we would have sampled only $23\%$ of the total $U$-band light from the clusters, and EL galaxies much fainter than our magnitude limit would constitute the dominant source of the EL flux. If the cluster GLF is indeed flatter than our EL GLF power law slope, as suggested by the results of \citet{trentham02}, then the real fraction of the light that we have sampled is likely to be near the top end of this range.

Are these results valid for any individual cluster? The numbers above apply to an ``average'' of the three clusters in our sample. However, we have demonstrated in \S 3.1 that, even if the GLFs in our clusters are consistent with each other in shape, they differ in normalization. To estimate the cumulative $L_{U}$ for any individual cluster, we therefore have to consider them separately. We use the composite GLF to constrain the shape parameters for the extrapolation under the assumption (which is consistent with the data) that the same fit describes all three clusters. We then renormalize the Schechter function to reproduce the total luminosities at $M_{U}<-18.5$ in each individual cluster. Table \ref{tabculums} shows the luminosities for the composite GLF and the individual clusters. It gives both the total luminosities at the bright end ($M_{U}<-18.5$) and the extrapolated luminosities to a limit of $M_{U}=-10$. The table shows that the systematic uncertainties about the shape of the faint end GLF, particularly whether there is an upturn or not, are larger than the random uncertainties associated with either extrapolation method.

Note that we have adopted a limit of $M_{U}=-18.5$ instead of the $M_{U}=-17.5$ that we regard as our confidence limit for the composite GLF. This is because, for the same reasons laid out in \S \ref{coverage}, the more distant clusters, A754 and A85, are only sampled to $M_{R}\approx-19.0$. For $M_{U}>-18.5$, significant numbers of galaxies could be lost at fainter $M_{R}$. However, we still use the full extent of the composite GLF (to $M_{U}=-17.5$) to constrain the Schechter parameters for the extrapolation.

\label{apptotallum}

\section{HOW MUCH LIGHT FROM LARGER RADII?}

\label{applargeradii}

What is the effect of the limited spatial coverage of our survey? We calculate galaxy number and luminosity profiles as a function of radius (normalized so that the harmonic radius is exactly 1 in each cluster). We then fit these profiles with isothermal beta models \citep{cavaliere76,cavaliere78} projected onto the radial dimension, and extrapolate the models to estimate the total number of galaxies and total $U$-band luminosity out to 4.1 harmonic radii ($\sim5$ Mpc in the case of A496, roughly the infall radius). This extrapolation is justified by recent results from \citet{gomez}, who show that star formation rates typically exhibit a break around 3-4 virial radii, comparable to the outer limit of our extrapolation and to the edge of the cluster.

\begin{figure}
\plotone{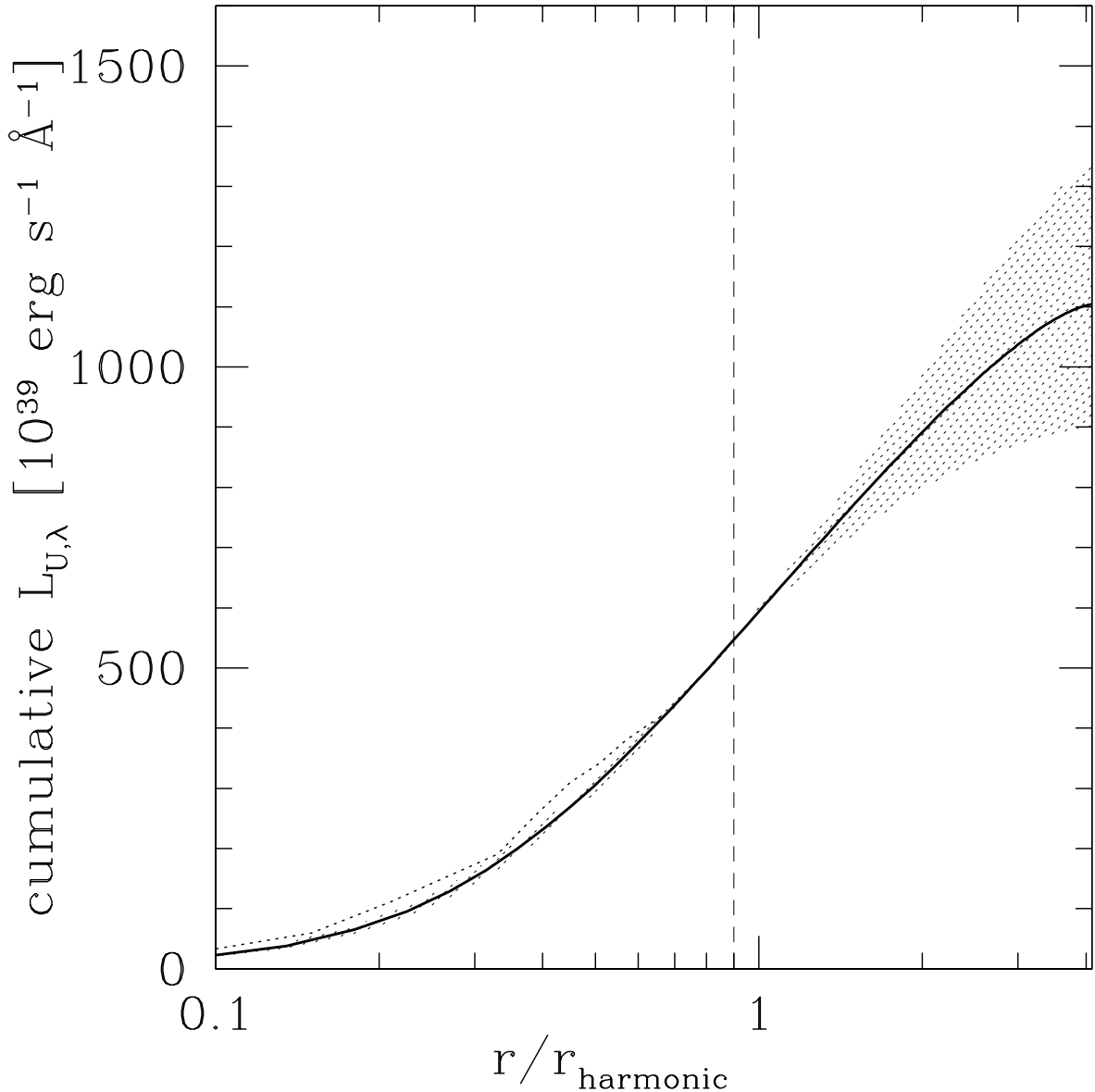}
\caption{Integrated $U$-band luminosity as a function of the cluster sampling radius in units of the harmonic radius for each cluster. The dotted line shows the measured profile out to a limit of $r=0.9r_{vir}$ (denoted by a thin vertical line). The dark solid line shows the extrapolation from the best fit isothermal-$\beta$ model. Errors are based on the 1$\sigma$ uncertainties in the best fit. The extrapolated profile is normalized to yield the correct sampled luminosity within $r=0.9r_{vir}$. Within this limit, we sample $\sim$50\% of the $U$-band light from galaxies with $M_{U}<-17$.}
\end{figure}

We estimate that our survey samples $0.50^{+0.11}_{-0.09}$ of the total $U$-band luminosity of a cluster from galaxies with $M_{U}<-17.5$ within the survey region of A496. The fraction of galaxies sampled in this region is $0.80^{+0.5}_{-0.6}$. This result indicates that the $U$-band luminosity profile is more extended than the galaxy number count profile, presumably due to radial color gradients in the cluster. Within 0.9 harmonic radii, our survey samples more than three quarters of the galaxies in our clusters, and about $1/2$ of the $U$-band luminosity from galaxies with $M_{U}=-17.5$.

\end{document}